\documentclass[prb,showpacs,preprint, preprintnumbers]{revtex4}%
\usepackage{amsfonts}
\usepackage{amsmath}
\usepackage{amssymb}
\usepackage{graphicx}%
\providecommand{\U}[1]{\protect\rule{.1in}{.1in}}

\begin{document}
\title{An extension of the Kubo-Greenwood formula for use in molecular simulations}
\author{Ming-Liang Zhang}
\affiliation{Department of Physics and Astronomy, Ohio University, Athens, Ohio 45701}
\author{D. A. Drabold}
\affiliation{Trinity College, Cambridge, CB2 1TQ, United Kingdom and Department of Physics
and Astronomy, Ohio University, Athens, Ohio 45701}

\begin{abstract}
We discuss the foundations and extend the range of applicability of the widely
used Kubo-Greenwood formula (KGF) for the electronic conductivity. The
conductivity is derived from the current density, and only the probability
amplitude rather than the transition probability is used. It is shown that the
contribution to the conductivity from degenerate states in a low or zero
frequency external electric field and the contribution from states near
resonance with a finite frequency external field are finite. The improved
conductivity expression does not include the familiar ``energy conserving"
delta function, and no artificial broadening parameter for delta function is
required for the DC limit. We explored two methods of computing current
density. We discuss the role of many-electron statistics in computing the
conductivity in single-particle approximations, and we show that the
conventional KGF is due to the contribution from single-particle excited states.


\end{abstract}

\maketitle

\section{Introduction}

The Kubo-Greenwood formula (KGF) has been widely used with great success to
calculate the electrical conductivity from first principles
simulations\cite{ale,gali,abt,heli,low,cle}. Despite the universal use of KGF,
it is worth pointing out that it has certain limitations, and for some
applications, improvements are possible. The aim of this paper is both to
derive the KGF and more fundamental transport formulae from first principles,
and to point out cases in which use of the KGF can be problematic, with a full
explanation of the origins of the difficulty. New formulae are presented which
circumvent some of these difficult cases.

Greenwood's derivation of the conductivity used the transition probability
between two single-electron states in an oscillating external field, in such a
way that the interaction time must be long enough to assure that the
transition probability is well-defined. On the other hand, to make
perturbation theory applicable, the interaction time should be
short\cite{Gre,md,mos,Over}. For a large system, in which the energy spectrum
is continuous, these two conditions are in conflict. The usual time-dependent
perturbation theory cannot be applied in two cases: (1) degenerate states and
(2) if two groups of states are in resonance with an oscillating external
field. Both circumstances are common in a macroscopic system. In addition, the
energy conserving $\delta$ functions in KGF require an artificial broadening
parameter when implemented numerically.

To avoid these difficulties, in Section II, both direct current (dc)
conductivity and alternating current (ac) conductivity are derived from a new
expression for the current density. Only the probability \textit{amplitude}
(not the probability itself) enters, and the $\delta$-function in the KGF
disappears in the improved expression of conductivity. The new expression may
be reduced to the well-known result for the conductivity by applying the
Boltzmann equation with the relaxation-time approximation for a crystal. For a
static field or oscillating field, the zeroth-order contribution from a group
of degenerate states is shown to vanish, and the first-order contribution of
degenerate states is finite (Appendix A). We also show that the zeroth-order
contribution from two groups of states which are near or in resonance with an
oscillating external field is zero. The first-order contribution of the two
groups of resonant states is finite. (Appendix B). By invoking the
single-particle approximation at different stages of the derivation, one is
led to slightly different results. In section III, we use many-body
perturbation theory to compute the conductivity for an intrinsic
semiconductor, metal and doped semiconductor. It is shown that at T=0K, the dc
conductivity of an intrinsic semiconductor is zero, a well-known consequence
of many-electron statistics.

\section{The Current Density}

If the motions of nuclei are treated classically (as in most \textit{ab
initio} MD codes), the average energy $\overline{H}$ of the electron + nuclei
system in an electromagnetic field described by vector potential $\mathbf{A}$
and scalar potential $\varphi$ is given by%
\begin{equation}
\overline{H}=\int d\mathbf{r}_{1}\cdots d\mathbf{r}_{N_{e}}\Lambda
_{\{\mathbf{W}_{1}\cdots\mathbf{W}_{\mathcal{N}}\}}^{\prime\ast}%
(\mathbf{r}_{1}\cdots\mathbf{r}_{N_{e}})H_{e}^{\prime}\Lambda_{\{\mathbf{W}%
_{1}\cdots\mathbf{W}_{\mathcal{N}}\}}^{\prime}(\mathbf{r}_{1}\cdots
\mathbf{r}_{N_{e}}) \label{enav}%
\end{equation}%
\[
+\sum_{\alpha}\frac{1}{2M_{\alpha}}(\mathbf{P}_{\alpha}-q_{\alpha}%
\mathbf{A}(\mathbf{W}_{\alpha}))^{2}+\sum_{\alpha}q_{\alpha}\varphi
(\mathbf{W}_{\alpha})+\sum_{\alpha,\beta(>\alpha)}V(\mathbf{W}_{\alpha
},\mathbf{W}_{\beta})
\]
where $q_{\alpha}$, $M_{\alpha},$ $\mathbf{P}_{\alpha}$ and $\mathbf{W}%
_{\alpha}$ are the effective charge, mass, canonical momentum and position
vector of the $\alpha^{\text{th}}$ nucleus. $V(\mathbf{W}_{\alpha}%
,\mathbf{W}_{\beta})$ is the interaction between the $\alpha^{\text{th}}$
nucleus and the $\beta^{\text{th}}$ nucleus.%
\begin{equation}
H_{e}^{\prime}=\sum_{j=1}^{N_{e}}[\frac{1}{2m}(\mathbf{p}_{j}-e\mathbf{A}%
(\mathbf{r}_{j}))^{2}+e\varphi(\mathbf{r}_{j})]+\sum_{jk}V(\mathbf{r}%
_{j},\mathbf{r}_{k})+\sum_{j\alpha}V(\mathbf{r}_{j},\mathbf{W}_{\alpha})
\label{dih}%
\end{equation}
is the electronic Hamiltonian in the external electromagnetic field.
$V(\mathbf{r}_{j},\mathbf{r}_{k})$ is the interaction between an electron at
$\mathbf{r}_{j}$ and another electron at $\mathbf{r}_{k}$, $V(\mathbf{r}%
_{j},\mathbf{W}_{\alpha})$ is the interaction between an eletron at
$\mathbf{r}_{j}$ and the $\alpha^{\text{th}}$ nucleus at $\mathbf{W}_{\alpha}%
$. The motion of the electrons is determined by%
\begin{equation}
H_{e}^{\prime}\Lambda_{\{\mathbf{W}_{1}\cdots\mathbf{W}_{\mathcal{N}}%
\}}^{\prime}(\mathbf{r}_{1}\cdots\mathbf{r}_{N_{e}})=E_{_{\{\mathbf{W}%
_{1}\cdots\mathbf{W}_{\mathcal{N}}\}}}^{\prime e}\Lambda_{\{\mathbf{W}%
_{1}\cdots\mathbf{W}_{\mathcal{N}}\}}^{\prime}(\mathbf{r}_{1}\cdots
\mathbf{r}_{N_{e}}) \label{ems}%
\end{equation}
$\Lambda_{\{\mathbf{W}_{1}\cdots\mathbf{W}_{\mathcal{N}}\}}^{\prime
}(\mathbf{r}_{1}\cdots\mathbf{r}_{N_{e}})$ is the many-electron wave function
of $H_{e}^{\prime}$ for a given nuclear configuration $\{\mathbf{W}_{1}%
\cdots\mathbf{W}_{\mathcal{N}}\}$ belonging to eigenvalue $E_{_{\{\mathbf{W}%
_{1}\cdots\mathbf{W}_{\mathcal{N}}\}}}^{\prime e}$. We use $H_{e}$ to denote
$H_{e}^{\prime}$ when external field does not appear, $\Lambda_{\{\mathbf{W}%
_{1}\cdots\mathbf{W}_{\mathcal{N}}\}}$ is the eigenfunction of $H_{e}$
belonging to eigenvalue $E_{_{\{\mathbf{W}_{1}\cdots\mathbf{W}_{\mathcal{N}%
}\}}}^{e}$. Hereafter we use a symbol with prime to denote a quantity when
external field appears, the corresponding symbol without prime to denote the
quantity in zero field. The velocity of the $\gamma^{\text{th}}$ nucleus is
determined by
\begin{equation}
\overset{\cdot}{\mathbf{W}}_{\gamma}=\frac{\partial\overline{H}}%
{\partial\mathbf{P}_{\gamma}}=\frac{\mathbf{P}_{\gamma}^{mech}}{M_{\gamma}%
}=\mathbf{V}_{\gamma} \label{vel}%
\end{equation}
where $\mathbf{V}_{\alpha}=\mathbf{P}_{\alpha}^{mech}/M_{\alpha}$ is the
velocity of the $\alpha^{\text{th}}$ nucleus, $\mathbf{P}_{\alpha}%
^{mech}=\mathbf{P}_{\alpha}-q_{\alpha}\mathbf{A}(\mathbf{W}_{\alpha})$ is the
mechanical momentum of the $\alpha^{\text{th}}$ nucleus. After some
manipulations, $\overset{\cdot}{\mathbf{P}}_{\gamma}=-\frac{\partial
\overline{H}}{\partial\mathbf{W}_{\gamma}}$ is read as%
\begin{equation}
M_{\gamma}\overset{\cdot}{\mathbf{V}}_{\gamma}=q_{\gamma}[\mathbf{E}%
(\mathbf{W}_{\gamma})+\mathbf{V}_{\gamma}\times\mathbf{B}(\mathbf{W}_{\gamma
}))-\sum_{\alpha(\neq\gamma)}\frac{\partial V(\mathbf{W}_{\alpha}%
,\mathbf{W}_{\gamma})}{\partial\mathbf{W}_{\gamma}}-\frac{\partial
E_{_{\{\mathbf{W}_{1}\cdots\mathbf{W}_{\mathcal{N}}\}}}^{\prime e}}%
{\partial\mathbf{W}_{\gamma}} \label{force}%
\end{equation}
In the MD formulation, the positions of the nuclei are functions of \ `time'
(MD\ steps). The initial positions of nuclei are given from an initial
configuration, the initial velocities of nuclei are assigned in some way. The
electronic wave function $\Lambda_{\{\mathbf{W}_{1}\cdots\mathbf{W}%
_{\mathcal{N}}\}}(\mathbf{r}_{1}\mathbf{r}_{2}\mathbf{r}_{3}\cdots
\mathbf{r}_{N_{e}})$ is calculated from the configuration $\{\mathbf{W}%
_{1}\cdots\mathbf{W}_{\mathcal{N}}\}$, the forces on each nucleus is then
calculated from $\Lambda_{\{\mathbf{W}_{1}\cdots\mathbf{W}_{\mathcal{N}}%
\}}(\mathbf{r}_{1}\mathbf{r}_{2}\mathbf{r}_{3}\cdots\mathbf{r}_{N_{e}})$. The
position and velocity of a nucleus in next\ step are calculated from the
length of the time-step, acceleration and the velocity in last step\cite{san}.

According to the principle of virtual work, for a given state $\Lambda
^{\prime}$, the microscopic electric current density $\mathbf{j}%
_{m}(\mathbf{r})$ at point $\mathbf{r}$ is\cite{ll}%
\begin{equation}
\mathbf{j}_{m}(\mathbf{r})=-\frac{\delta\overline{H}}{\delta\mathbf{A}%
(\mathbf{r})} \label{mdliu}%
\end{equation}%
\[
=N_{e}\frac{i\hbar e}{2m}\int d\mathbf{r}_{2}d\mathbf{r}_{3}\cdots
d\mathbf{r}_{N_{e}}(\Lambda^{\prime}\nabla_{\mathbf{r}_{1}}\Lambda^{\prime
\ast}-\Lambda^{\prime\ast}\nabla_{\mathbf{r}_{1}}\Lambda^{\prime})-\frac
{e^{2}}{m}\mathbf{A}(\mathbf{r})n_{\{\mathbf{W}_{1}\cdots\mathbf{W}%
_{\mathcal{N}}\}}^{e}(\mathbf{r})+\sum_{\alpha}q_{\alpha}\mathbf{V}_{\alpha
}\delta(\mathbf{r}-\mathbf{W}_{\alpha})
\]
where%
\begin{equation}
n_{\{\mathbf{W}_{1}\cdots\mathbf{W}_{\mathcal{N}}\}}^{e}(\mathbf{r}%
,t)=N_{e}\int d\mathbf{r}_{2}d\mathbf{r}_{3}\cdots d\mathbf{r}_{N_{e}}%
\Lambda_{\{\mathbf{W}_{1}\cdots\mathbf{W}_{\mathcal{N}}\}}^{\prime\ast
}(\mathbf{rr}_{2}\mathbf{r}_{3}\cdots\mathbf{r}_{N_{e}})\Lambda_{\{\mathbf{W}%
_{1}\cdots\mathbf{W}_{\mathcal{N}}\}}^{\prime}(\mathbf{rr}_{2}\mathbf{r}%
_{3}\cdots\mathbf{r}_{N_{e}}) \label{dene}%
\end{equation}
is the number density of electrons at $\mathbf{r}$ for a given nuclear
configuration $\{\mathbf{W}_{1}\cdots\mathbf{W}_{\mathcal{N}}\}$.
Eq.(\ref{mdliu}) is the response of the electrons+nuclei system to the
external field; the first two terms are due to electrons, and the last term is
due to nuclei. The measured macroscopic current density at point $\mathbf{r}%
$\ is\cite{robin,jac} a spatial average of Eq.(\ref{mdliu}) over a region
$\Omega_{\mathbf{r}}$ centered at $\mathbf{r}$:%
\begin{equation}
\mathbf{j}(\mathbf{r})=\frac{1}{\Omega_{\mathbf{r}}}\int_{\Omega_{\mathbf{r}}%
}d\mathbf{sj}_{m}(\mathbf{s}) \label{sap}%
\end{equation}
The linear size $L$ of $\Omega_{\mathbf{r}}$ satisfies: $a<<L<<\lambda$, where
$a$ is a typical bond length, $\lambda$ is the wavelength of external field or
other macroscopic length scale. Eq.(\ref{sap}) is the usual current density
defined for an infinitesimal area\cite{robin,jac}.

Using the single-electron approximation to separate variables in
Eq.(\ref{ems}), we obtain the equation satisfied by the single-electron wave
function $\chi_{l}^{\prime}$:%
\begin{equation}
h_{a}^{\prime}\chi_{l}^{\prime}(\mathbf{r})=E_{l}^{\prime}\chi_{l}^{\prime
}(\mathbf{r}),\text{ \ }h_{a}^{\prime}=\frac{1}{2m}(\mathbf{p}-e\mathbf{A}%
(\mathbf{r}))^{2}+e\varphi(\mathbf{r})+U(\mathbf{r},\{\mathbf{W}_{\alpha}\})
\label{sha}%
\end{equation}
where $h_{a}^{\prime}$ is the single-electron Hamiltonian in an external
field, $U$ is the single-electron potential due to nuclear configuration
$\{\mathbf{W}_{\alpha}\}$. $h_{a}$, $\chi_{l}(\mathbf{r})$ and $E_{l}$ are the
corresponding quantities when external field does not appear. They are the
Hamiltonian, eigenfunctions and eigenvalues as in density functional theory
(DFT), or other single particle theories.

The current density due to electrons can be computed as following. At finite
temperature $T>0$, the system can be in the ground or excited states. The
electron current at temperature T comes from both the various excited states
and the ground state:%
\begin{equation}
\mathbf{j}^{e}(\mathbf{r})=\frac{i\hbar eN_{e}}{2m\Omega_{\mathbf{r}}}%
\int_{\Omega_{\mathbf{r}}}d\mathbf{s}\int d\mathbf{r}_{2}d\mathbf{r}_{3}\cdots
d\mathbf{r}_{N_{e}}\sum_{l_{1}l_{2}\cdots l_{N_{e}}}W^{\prime}{}_{l_{1}%
l_{2}\cdots l_{N_{e}}}(\Lambda_{l_{1}l_{2}\cdots l_{N_{e}}}^{\prime}%
\nabla_{\mathbf{s}}\Lambda_{l_{1}l_{2}\cdots l_{N_{e}}}^{\prime\ast}%
-\Lambda_{l_{1}l_{2}\cdots l_{N_{e}}}^{\prime\ast}\nabla_{\mathbf{s}}%
\Lambda_{l_{1}l_{2}\cdots l_{N_{e}}}^{\prime}) \label{liup}%
\end{equation}
where%
\begin{equation}
\Lambda_{l_{1}l_{2}\cdots l_{N_{e}}}^{\prime}=\frac{1}{\sqrt{N_{e}!}}\sum
_{P}\delta_{P}P\chi_{l_{1}}^{\prime}(\mathbf{s},s_{z1})\chi_{l_{2}}^{\prime
}(\mathbf{r}_{2},s_{z2})\chi_{l_{3}}^{\prime}(\mathbf{r}_{3},s_{z3})\cdots
\chi_{l_{N_{e}}}^{\prime}(\mathbf{r}_{N_{e}},s_{zN_{e}}) \label{sla}%
\end{equation}
is a $N_{e}-$electron state, $P$ is a permutation on $N_{e}$ objects
($\mathbf{r}_{1}s_{z1};\mathbf{r}_{2}s_{z2};\mathbf{r}_{3}s_{z3}%
;\cdots;\mathbf{r}_{N_{e}}s_{zN_{e}}$), $\delta_{P}=1$ if $P$ is an even
permutation, $\delta_{P}=-1$ if P is odd. Of course $l_{1},l_{2}%
,\cdots,l_{N_{e}}$ are distinct. Because any observable like $\mathbf{j}^{e}$
is bilinear about $\Lambda_{l_{1}l_{2}\cdots l_{N_{e}}}^{\prime}$, the order
of rows and the order of columns in $\Lambda_{l_{1}l_{2}\cdots l_{N_{e}}%
}^{\prime}$ do not matter. We only need to maintain a fixed order in all
intermediate steps of calculation. The sum is over all possible choices of
$N_{e}$ single-electron states. The arguments of $\Lambda^{\prime}$ are
$(\mathbf{s},\mathbf{r}_{2},\mathbf{r}_{3},\cdots,\mathbf{r}_{N_{e}})$, to
save space the spin variables are abbreviated.%
\begin{equation}
W_{l_{1}l_{2}\cdots l_{N_{e}}}^{\prime}=U_{l_{1}l_{2}\cdots l_{N_{e}}}%
^{\prime}/Z^{\prime},\text{ \ }Z^{\prime}=\sum_{l_{1}l_{2}\cdots l_{N_{e}}%
}U_{l_{1}l_{2}\cdots l_{N_{e}}}^{\prime},\text{ }U_{l_{1}l_{2}\cdots l_{N_{e}%
}}^{\prime}=\exp[-(E_{l_{1}l_{2}\cdots l_{N_{e}}}^{\prime}-E_{0}^{\prime
})/(k_{B}T)] \label{abp}%
\end{equation}
is the appearing probability of state $\Lambda_{l_{1}l_{2}\cdots l_{N_{e}}%
}^{\prime}$. $E_{0}^{\prime}$ is the energy of $N_{e}-$electron ground state.
When no field is applied on the system, macroscopic current does not appear in
any state $\Lambda_{l_{1}l_{2}\cdots l_{N_{e}}}$. The current density from
electrons reads:%
\[
\mathbf{j}^{e}(\mathbf{r})=\frac{i\hbar eN_{e}}{2m\Omega_{\mathbf{r}}}%
\int_{\Omega_{\mathbf{r}}}d\mathbf{s}\int d\mathbf{r}_{2}d\mathbf{r}_{3}\cdots
d\mathbf{r}_{N_{e}}\sum_{l_{1}l_{2}\cdots l_{N_{e}}}W_{l_{1}l_{2}\cdots
l_{N_{e}}}^{\prime}[(\Lambda_{l_{1}l_{2}\cdots l_{N_{e}}}^{\prime}\nabla
_{s}\Lambda_{l_{1}l_{2}\cdots l_{N_{e}}}^{\prime\ast}-\Lambda_{l_{1}%
l_{2}\cdots l_{N_{e}}}^{\prime\ast}\nabla_{s}\Lambda_{l_{1}l_{2}\cdots
l_{N_{e}}}^{\prime})
\]%
\begin{equation}
-(\Lambda_{l_{1}l_{2}\cdots l_{N_{e}}}\nabla_{s}\Lambda_{l_{1}l_{2}\cdots
l_{N_{e}}}^{\ast}-\Lambda_{l_{1}l_{2}\cdots l_{N_{e}}}^{\ast}\nabla_{s}%
\Lambda_{l_{1}l_{2}\cdots l_{N_{e}}})](\mathbf{sr}_{2}\mathbf{r}_{3}%
\cdots\mathbf{r}_{N_{e}}) \label{cru}%
\end{equation}

For low temperatures, $\langle\chi_{l_{1}}|-e\mathbf{E}\cdot\mathbf{r}%
|\chi_{l_{1}}\rangle<<k_{B}T$ is not satisfied. Linearizing $W_{l_{1}%
l_{2}\cdots l_{N_{e}}}^{\prime}$ about field $\mathbf{E}$ is not legitimate
(cf. Eq. (\ref{abp})): current density is not necessary linear about field,
the dependence of conductivity on field is intrinsic at low temperature for
semiconductors. If temperature is \textit{not too low} ($\langle\chi_{l_{1}%
}|-eEr|\chi_{l_{1}}\rangle<<k_{B}T$), we may expand $W^{\prime}[]$ in
Eq.(\ref{cru}) to first order of field%
\begin{equation}
W^{\prime}[]=W[]_{E=0}+[]_{E=0}\sum_{\alpha}E_{\alpha}\frac{\partial
W}{\partial E_{\alpha}}+W\sum_{\alpha}E_{\alpha}\frac{\partial\lbrack
]}{\partial E_{\alpha}}=W\sum_{\alpha}E_{\alpha}\frac{\partial\lbrack
]}{\partial E_{\alpha}} \label{xian}%
\end{equation}
The last equal sign used the obvious fact $[]_{E=0}=0$: no macroscopic current
exist when external field vanishes. With the help of Eq.(\ref{xian}),
Eq.(\ref{cru}) is simplified to%
\[
\mathbf{j}^{e}(\mathbf{r})=\frac{i\hbar eN_{e}}{2m\Omega_{\mathbf{r}}}%
\int_{\Omega_{\mathbf{r}}}d\mathbf{s}\int d\mathbf{r}_{2}d\mathbf{r}_{3}\cdots
d\mathbf{r}_{N_{e}}\sum_{l_{1}l_{2}\cdots l_{N_{e}}}W_{l_{1}l_{2}\cdots
l_{N_{e}}}[(\Lambda_{l_{1}l_{2}\cdots l_{N_{e}}}^{\prime}\nabla_{\mathbf{s}%
}\Lambda_{l_{1}l_{2}\cdots l_{N_{e}}}^{\prime\ast}-\Lambda_{l_{1}l_{2}\cdots
l_{N_{e}}}^{\prime\ast}\nabla_{\mathbf{s}}\Lambda_{l_{1}l_{2}\cdots l_{N_{e}}%
}^{\prime})
\]%
\begin{equation}
-(\Lambda_{l_{1}l_{2}\cdots l_{N_{e}}}\nabla_{\mathbf{s}}\Lambda_{l_{1}%
l_{2}\cdots l_{N_{e}}}^{\ast}-\Lambda_{l_{1}l_{2}\cdots l_{N_{e}}}^{\ast
}\nabla_{\mathbf{s}}\Lambda_{l_{1}l_{2}\cdots l_{N_{e}}})](\mathbf{sr}%
_{2}\mathbf{r}_{3}\cdots\mathbf{r}_{N_{e}}) \label{urc}%
\end{equation}
where%
\begin{equation}
W_{l_{1}l_{2}\cdots l_{N_{e}}}=U_{l_{1}l_{2}\cdots l_{N_{e}}}/Z,\text{
\ }Z=\sum_{l_{1}l_{2}\cdots l_{N_{e}}}U_{l_{1}l_{2}\cdots l_{N_{e}}},\text{
\ }U_{l_{1}l_{2}\cdots l_{N_{e}}}=\exp[-(E_{l_{1}l_{2}\cdots l_{N_{e}}%
}-E_{v_{1}v_{2}\cdots v_{N_{e}}})/(k_{B}T)] \label{pro}%
\end{equation}
are the corresponding quantities without external field. In the
single-particle approximation%
\begin{equation}
W_{l_{1}l_{2}\cdots l_{N_{e}}}=%
{\displaystyle\prod\limits_{\alpha=1}^{N_{e}}}
f(E_{l_{\alpha}}),\text{ \ \ }f(E_{l_{\alpha}})=\frac{1}{e^{(E_{l_{\alpha}%
}-\mu)/k_{B}T}+1} \label{wps}%
\end{equation}
where $\mu$ is chemical potential at given temperature and shape of the
interested body.

The current density (\ref{urc}) and the conductivity deduced from it are just
for one MD step. To include the the thermal vibrations in a material, one must
average the conductivity over many MD steps. Only the averaged conductivity
may be compared to the experimental observations where the material changes
its configurations with time through thermal vibrations. This observation is
valid for solids, liquids and molecules.

The idea of linear response\cite{kub} can be applied in two different ways:
(1) first express $\Lambda_{l_{1}l_{2}\cdots l_{N_{e}}}^{\prime}$ with
single-electron wave functions $\chi^{\prime}$ and effect the multiple
integral $\int d\mathbf{r}_{2}d\mathbf{r}_{3}\cdots d\mathbf{r}_{N_{e}}$. Then
view $\chi_{l_{\alpha}}^{\prime}$ as correction of $\chi_{l_{\alpha}}$ under
perturbation $-e\mathbf{E}\cdot\mathbf{r}$. (2) view $\Lambda_{l_{1}%
l_{2}\cdots l_{N_{e}}}^{\prime}$ as correction of $\Lambda_{l_{1}l_{2}\cdots
l_{N_{e}}}$ under perturbation $-\sum_{m=1}^{N_{e}}e\mathbf{E}\cdot
\mathbf{r}_{m}$. Then effect multiple integral $\int d\mathbf{r}%
_{2}d\mathbf{r}_{3}\cdots d\mathbf{r}_{N_{e}}$. Most discussion about the
KGF\cite{Gre,md,mos,Over} is based on method (1). The role of many-electron
statistics is displayed more explicitly in method (2). Both schemes express
the conductivity in terms of single-electron states and corresponding
eigenvalues. In the remainder of this section, we will use method (1) and
compare with previous result. Method (2) will be analyzed in the next section.

\subsection{DC conductivity}

Applying Eq.(\ref{sla}) and working out $\int d\mathbf{r}_{2}d\mathbf{r}%
_{3}\cdots d\mathbf{r}_{N_{e}}$, Eq.(\ref{urc}) leads to:%
\begin{equation}
\mathbf{j}^{e}(\mathbf{r})=\frac{i\hbar e}{2m\Omega_{\mathbf{r}}}\int
_{\Omega_{\mathbf{r}}}d\mathbf{s}\sum_{l_{1}l_{2}\cdots l_{N_{e}}}%
W_{l_{1}l_{2}\cdots l_{N_{e}}}\sum_{\alpha=1}^{N_{e}}[(\chi_{l_{\alpha}%
}^{\prime}\nabla_{\mathbf{s}}\chi_{l_{\alpha}}^{\prime\ast}-\chi_{l_{\alpha}%
}^{\prime\ast}\nabla_{\mathbf{s}}\chi_{l_{\alpha}}^{\prime})-(\chi_{l_{\alpha
}}\nabla_{\mathbf{s}}\chi_{l_{\alpha}}^{\ast}-\chi_{l_{\alpha}}^{\ast}%
\nabla_{\mathbf{s}}\chi_{l_{\alpha}})] \label{sux}%
\end{equation}
the argument of all single-electron functions is $\mathbf{s}$.

In a static electric field, the nuclei and the bound electrons are pushed in
opposite directions. These lead to a static deformation of the material. Since
a static electric field does not produce any net velocities of nuclei, the 3rd
term in Eq.(\ref{mdliu}) is zero. A static electric field is solely determined
by scalar potential $\varphi(\mathbf{r})$, which means $\mathbf{A}%
(\mathbf{r})=0$. The 2nd term in Eq.(\ref{mdliu}) vanishes. The interaction
with an\ electron at $\mathbf{r}$ is
\begin{equation}
H_{fm}=e\varphi(\mathbf{r})=-e\mathbf{E}\cdot\mathbf{r} \label{sfm}%
\end{equation}
At this point, let us assume all the single-electron states in Eq.(\ref{sux})
are non-degenerate. The case of degenerate states will be discussed later.
From first-order perturbation theory, the change $\chi_{c}^{\prime(1)}$ in the
single electron wave function due to the external field is%
\begin{equation}
\chi_{c}^{\prime}=\chi_{c}+\chi_{c}^{\prime(1)},\text{ \ \ \ }\chi_{c}%
^{\prime(1)}=\sum_{d(\neq c)}\frac{\langle\chi_{d}|-e\mathbf{E}\cdot
\mathbf{r}|\chi_{c}\rangle}{E_{c}-E_{d}}\chi_{d} \label{1wf}%
\end{equation}
$\chi_{c}$ and $E_{c}$ are the single electron wave function and the
corresponding eigenvalue without external field. We should emphasize that
voltage is proportional to the distance between two points, and so too is the
interaction. The change in states cannot be described by the perturbation
result Eq.(\ref{1wf}). Except for very weak field, one must use WKB method
rather than perturbation theory. In this work, let us limit ourselves to very
weak field. Substituting Eq.(\ref{1wf}) into Eq.(\ref{sux}), and only keeping
the terms linear with external field, one has%
\begin{equation}
\mathbf{j}^{e}(\mathbf{r})=\frac{i\hbar e}{2m\Omega_{\mathbf{r}}}\int
_{\Omega_{\mathbf{r}}}d\mathbf{s}\sum_{l_{1}l_{2}\cdots l_{N_{e}}}%
W_{l_{1}l_{2}\cdots l_{N_{e}}}\sum_{\alpha=1}^{N_{e}}\sum_{l(\neq l_{\alpha}%
)}\frac{1}{E_{l_{\alpha}}-E_{l}} \label{usa}%
\end{equation}%
\[
\{\langle\chi_{l}|e\mathbf{E}\cdot\mathbf{r}|\chi_{l_{\alpha}}\rangle
(\chi_{l_{\alpha}}^{\ast}\nabla_{\mathbf{s}}\chi_{l}-\chi_{l}\nabla
_{\mathbf{s}}\chi_{l_{\alpha}}^{\ast})-\langle\chi_{l}|e\mathbf{E}%
\cdot\mathbf{r}|\chi_{l_{\alpha}}\rangle^{\ast}(\chi_{l_{\alpha}}^{\ast}%
\nabla_{\mathbf{s}}\chi_{l}-\chi_{l}\nabla_{\mathbf{s}}\chi_{l_{\alpha}}%
^{\ast})^{\ast}\}
\]
The sum over $l(\neq l_{\alpha})$ is not restrict to ($l_{1}l_{2}\cdots
l_{N_{e}}$); it extends to all single particle states. By means of the
definition of conductivity $\sigma_{\mu\nu}$%
\begin{equation}
j_{\mu}=\sum_{\nu}\sigma_{\mu\nu}E_{\nu},\text{ \ \ }\mu,\nu=x,y,z \label{cdu}%
\end{equation}
the dc conductivity is%
\begin{equation}
\sigma_{\mu\nu}=\frac{e^{2}\hbar}{m\Omega}\sum_{l_{1}l_{2}\cdots l_{N_{e}}%
}W_{l_{1}l_{2}\cdots l_{N_{e}}}\sum_{\alpha=1}^{N_{e}}\sum_{l(\neq l_{\alpha
})}\frac{1}{E_{l_{\alpha}}-E_{l}}\operatorname{Im}\langle\chi_{l}|x_{\nu}%
|\chi_{l_{\alpha}}\rangle\int_{\Omega}d^{3}x(\chi_{l}\frac{\partial
\chi_{l_{\alpha}}^{\ast}}{\partial x_{\mu}}-\chi_{l_{\alpha}}^{\ast}%
\frac{\partial\chi_{l}}{\partial x_{\mu}}) \label{stc}%
\end{equation}
In a large system, the matrix element of position operator is not well
defined. Making use of%
\begin{equation}
\langle\chi_{d}|x_{\alpha}|\chi_{l_{\alpha}}\rangle=\frac{\langle\chi
_{d}|[h_{a},x_{\alpha}]|\chi_{l_{\alpha}}\rangle}{E_{d}-E_{l_{\alpha}}}%
=\frac{\hbar^{2}}{m}\frac{\langle\chi_{d}|\frac{\partial}{\partial x_{\alpha}%
}|\chi_{l_{\alpha}}\rangle}{E_{l_{\alpha}}-E_{d}} \label{uid}%
\end{equation}
one can change the matrix element of position operator into the matrix element
of momentum operator\cite{md}.

Current use of the KGF is amounts to assuming that beside $l_{\alpha}$, other
single-electron states in $\{l_{1}l_{2}\cdots l_{N_{e}}\}$ are occupied. Only
the factor $f(E_{l_{\alpha}})$ is left. Thus the sum over various$\ $choices
of $\{l_{1}l_{2}\cdots l_{N_{e}}\}$ can be ignored if one extends the sum over
$\alpha$ to all possible single particle states.

In parallel with Greenwood's work for ac field, Luttinger has derived an
expression for static field by adiabatically introducing the
interaction.\cite{lut} Transition probability rather than the amplitude of
probability was used.\ Eq.(\ref{stc}) does not obviously display a feature of
an intrinsic semiconductor: dc conductivity vanishes at zero temperature. In
addition, due to the use of the single particle approximation before applying
perturbation theory in Eq.(\ref{stc}), one cannot exclude coupling between two
occupied states. These faults can be cured in time-dependent perturbation
theory or by applying perturbation theory directly to the many-electron wave function.

If there is only one group $M$ degenerate single-electron states
($\chi_{d_{\sigma}},\sigma=1,2,\cdots,M$) in $\Lambda_{l_{1},l_{2}%
,\cdots,l_{N_{e}}}$, we first form correct zeroth order wave functions%
\begin{equation}
\chi_{d_{\sigma}}^{\prime(0)}=\sum_{\sigma^{\prime}}C_{d_{\sigma}%
d_{\sigma^{\prime}}}\chi_{d_{\sigma^{\prime}}},\text{ \ }\sigma,\sigma
^{\prime}=1,2,\cdots,M \label{0wf}%
\end{equation}
the secular equation satisfied by $C_{d_{\sigma}d_{\sigma^{\prime}}}$ is%
\begin{equation}
\sum_{\sigma^{\prime}}(V_{d_{\sigma}d_{\sigma^{\prime}}}-\varepsilon
\delta_{d_{\sigma}d_{\sigma^{\prime}}})C_{d_{\sigma}d_{\sigma^{\prime}}%
}=0,\text{ \ \ }V_{d_{\sigma}d_{\sigma^{\prime}}}=\int d\mathbf{r}%
\chi_{d_{\sigma}}^{\ast}(-e\mathbf{E}\cdot\mathbf{r})\chi_{d_{\sigma^{\prime}%
}} \label{jiu}%
\end{equation}
The perturbation matrix ($V_{d_{\sigma}d_{\sigma^{\prime}}}$) is Hermitian, it
can be diagonalized by a unitary transformation ($C_{d_{\sigma}d_{\sigma
^{\prime}}}$). Therefore%
\begin{equation}
\sum_{\sigma}(\chi_{d_{\sigma}}^{\prime(0)}\nabla\chi_{d_{\sigma}}%
^{\prime(0)\ast}-\chi_{d_{\sigma}}^{\prime(0)\ast}\nabla\chi_{d_{\sigma}%
}^{\prime(0)})=\sum_{\sigma}(\chi_{d_{\sigma}}\nabla\chi_{d_{\sigma}}^{\ast
}-\chi_{d_{\sigma}}^{\ast}\nabla\chi_{d_{\sigma}}) \label{noac}%
\end{equation}
According to Eq.(\ref{sux}), the zeroth-order contribution of the degenerate
states to conductivity is zero. This is consistent with usual experience: an
electron is not accelerated along the direction of the field when it transits
between states with same energy, and thus makes no contribution to the conductivity.

The first order correction to $\chi_{d_{\sigma}}^{\prime(0)}$ is\cite{ll}%
\begin{equation}
\chi_{d_{\sigma}}^{\prime(1)}=\sum_{k}\frac{V_{kd_{\sigma}}}{E_{d_{\sigma}%
}^{(0)}-E_{k}^{(0)}}\chi_{k}+\sum_{\sigma^{\prime}=1}^{M}[\frac{1}%
{\varepsilon_{d_{\sigma}}-\varepsilon_{d_{\sigma^{\prime}}}}\sum_{k}%
\frac{V_{d_{\sigma^{\prime}}k}V_{kd_{\sigma}}}{E_{d_{\sigma}}^{(0)}%
-E_{k}^{(0)}}]\chi_{d_{\sigma^{\prime}}}^{\prime(0)} \label{1dg}%
\end{equation}
where%
\begin{equation}
\varepsilon_{d_{\sigma}}=\int d\mathbf{r}\chi_{d_{\sigma}}^{\prime\ast
}(-e\mathbf{E}\cdot\mathbf{r})\chi_{d_{\sigma}}^{\prime},\text{ \ \ }%
V_{kd_{\sigma}}=\int d\mathbf{r}\chi_{k}^{\ast}(-e\mathbf{E}\cdot
\mathbf{r})\chi_{d_{\sigma}}^{\prime(0)},\text{ \ }V_{d_{\sigma^{\prime}}%
k}=\int d\mathbf{r}\chi_{d_{\sigma^{\prime}}}^{\prime(0)\ast}(-e\mathbf{E}%
\cdot\mathbf{r})\chi_{k} \label{mdg}%
\end{equation}
$k$ indexes non-degenerate states. The first sum in Eq.(\ref{1dg}) runs over
all states which are not degenerate with ($\chi_{d_{\sigma}},\sigma
=1,2,\cdots,M$).

Making use of Eq.(\ref{1dg}), the macroscopic current density Eq.(\ref{sux})
becomes%
\[
\mathbf{j}(\mathbf{r}^{\prime})=\frac{e^{2}\hbar}{m\Omega_{\mathbf{r}^{\prime
}}}\sum_{l_{1}l_{2}\cdots l_{N_{e}}}W_{l_{1}l_{2}\cdots l_{N_{e}}}%
\{\sum_{\alpha=1}^{N_{e}-M}\sum_{l(\neq l_{\alpha})}\operatorname{Im}%
\frac{\langle\chi_{l}|\mathbf{E}\cdot\mathbf{r}|\chi_{l_{\alpha}}\rangle
}{E_{l_{\alpha}}-E_{l}}\int_{\Omega_{\mathbf{r}^{\prime}}}d\mathbf{s}(\chi
_{l}\nabla_{\mathbf{s}}\chi_{l_{\alpha}}^{\ast}-\chi_{l_{\alpha}}^{\ast}%
\nabla_{\mathbf{s}}\chi_{l})
\]%
\begin{equation}
+\sum_{\sigma=1}^{M}\sum_{k}\operatorname{Im}\frac{\langle\chi_{k}%
|\mathbf{E}\cdot\mathbf{r}|\chi_{d_{\sigma}}\rangle}{E_{d_{\sigma}}%
^{(0)}-E_{k}^{(0)}}\int_{\Omega_{\mathbf{r}^{\prime}}}d\mathbf{s}(\chi
_{k}\nabla_{\mathbf{s}}\chi_{d_{\sigma}}^{\prime(0)\ast}-\chi_{d_{\sigma}%
}^{\prime(0)\ast}\nabla_{\mathbf{s}}\chi_{k}) \label{uds}%
\end{equation}%
\[
+\sum_{\sigma=1}^{M}\sum_{\sigma^{\prime}(\neq\sigma)}\operatorname{Im}%
[\frac{1}{\varepsilon_{d_{\sigma}}-\varepsilon_{d_{\sigma^{\prime}}}}\sum
_{k}\frac{V_{d_{\sigma^{\prime}}k}V_{kd_{\sigma}}}{E_{d_{\sigma}}^{(0)}%
-E_{k}^{(0)}}]\int_{\Omega_{\mathbf{r}^{\prime}}}d\mathbf{s}(\chi_{d_{\sigma}%
}^{\prime(0)\ast}\nabla_{\mathbf{s}}\chi_{d_{\sigma^{\prime}}}^{\prime
(0)}-\chi_{d_{\sigma^{\prime}}}^{\prime(0)}\nabla_{\mathbf{s}}\chi_{d_{\sigma
}}^{\prime(0)\ast})\}
\]
The sums over $l$ and $k$ are not restrict to ($l_{1}l_{2}\cdots l_{N_{e}}$).
The first term in bracket is the contribution from coupling among
non-degenerate states, the third term is contribution from coupling among
M-fold degenerate states, the second term is contribution from coupling
between member of non-degenerate states and the M-fold degenerate states. All
denominators are non-zero.

Because $\chi_{d_{\sigma}}^{\prime(0)}$, $\varepsilon_{d_{\sigma}}$ and
$V_{kd_{\sigma}}$ are functions of field, the 2$^{\text{nd}}$ term and the
3$^{\text{rd}}$ term in Eq.(\ref{uds}) do not exhibit a simple linear relation
with field: each contribution to conductivity is field-dependent. The
generalization to the situation in which there are several groups of
degenerate states in $\Lambda_{l_{1},l_{2},\cdots,l_{N_{e}}}$ is straightforward.

The massive degeneracies in the $N_{e}$\textit{-electron states }%
$\Lambda_{l_{1},l_{2},\cdots,l_{N_{e}}}$ do not cause any trouble: they are
counted by the sum over all possible ways of picking up $N_{e}$
single-electron states. In a crystal, the degeneracies in the
\textit{single-electron states} may be caused by the high symmetry of the
lattice. Sometimes accidental degeneracies (band crossing along a symmetry
axis or a symmetry plane not compelled \ by symmetry) also occur\cite{her}),
as well as additional degeneracies produced by time reversal
symmetry\cite{ott,cov}. In a disordered system (liquid, amorphous solid and
some molecules), only the degeneracies produced by time reversal symmetry are
left. From the perspective of computing the conductivity, a disordered system
has less degeneracy, and is simpler to treat than a crystal. Comparing the
huge number of non-degenerate states (most of original degeneracies in the
unit cell are removed by the interaction with first, second and third
coordination shells), the number of degenerate states in each manifold is
quite small. The third term in Eq.(\ref{uds}), the coupling among degenerate
states, is small compared with the first two terms. If we neglect the
field-dependence of $\chi_{d_{\sigma}}^{\prime(0)}$ in the second term of
Eq.(\ref{uds}) and use the original $\chi_{d_{\sigma}}$, one does not need to
distinguish degenerate states and non-degenerate states. Eq.(\ref{stc}) could
be used to compute conductivity by neglecting the coupling among degenerate states.

\subsection{AC conductivity}

The macroscopic current density in an oscillating field is%

\begin{equation}
\mathbf{j}^{e}(\mathbf{r},t)=\frac{i\hbar e}{2m\Omega_{\mathbf{r}}}%
\int_{\Omega_{\mathbf{r}}}d\mathbf{s}\sum_{l_{1}l_{2}\cdots l_{N_{e}}}%
W_{l_{1}l_{2}\cdots l_{N_{e}}}\sum_{\alpha=1}^{N_{e}} \label{acc}%
\end{equation}%
\[
\{[\chi_{l_{\alpha}}(t)\nabla_{\mathbf{s}}\chi_{l_{\alpha}}^{\prime(1)\ast
}(t)-\chi_{l_{\alpha}}^{\ast}(t)\nabla_{\mathbf{s}}\chi_{l_{\alpha}}%
^{\prime(1)}(t)]+[\chi_{l_{\alpha}}^{\prime(1)}(t)\nabla_{\mathbf{s}}%
\chi_{l_{\alpha}}^{\ast}(t)-\chi_{l_{\alpha}}^{\prime(1)\ast}(t)\nabla
_{\mathbf{s}}\chi_{l_{\alpha}}(t)]\}
\]
In an ac electric field $\mathbf{E}=\mathbf{E}_{0}\cos\omega t$, the
interaction of an electron at $\mathbf{r}$ with field is%
\begin{equation}
H_{fm}(t)=Fe^{-it\omega}+Fe^{it\omega},\text{ \ \ }F=-\frac{1}{2}%
e\mathbf{r\cdot E}_{0} \label{afm}%
\end{equation}
Since the region $\Omega_{\mathbf{r}}$ (employed to compute the spatial
average) is much smaller than the wavelength of the field, the position
dependence of field is ignored in Eq.(\ref{afm}). The wave function $\chi
_{c}^{\prime}(t)$ in an external field can be computed from time-dependent
perturbation theory%

\begin{equation}
\chi_{c}^{\prime}(t)=\chi_{c}(t)+\sum_{d(\neq c)}a_{d}(t)\chi_{d}%
e^{-itE_{d}/\hbar},\text{ \ }\chi_{c}(t)=e^{-itE_{c}/\hbar}\chi_{c}%
(\mathbf{r}) \label{tp1}%
\end{equation}
where $a_{d}(t)$ satisfies%
\begin{equation}
i\hbar\frac{\partial a_{d}(t)}{\partial t}=\sum_{c_{1}}a_{c_{1}}(t)F_{dc_{1}%
}[e^{it(\omega_{dc_{1}}-\omega)}+e^{it(\omega_{dc_{1}}+\omega)}] \label{coe1}%
\end{equation}
where%
\begin{equation}
F_{dc_{1}}=\int d\mathbf{r}\chi_{d}^{\ast}F\chi_{c_{1}}\text{ \ \ \ }%
\omega_{dc_{1}}=\frac{1}{\hbar}(E_{d}-E_{c_{1}}) \label{can}%
\end{equation}
We assume initially only state $\chi_{c}$ is occupied and other states are
empty: $a_{c_{1}}(t=-\infty)=\delta_{cc_{1}}$.

\subsubsection{non-degenerate states}

If all the states are not degenerate, Eq.(\ref{coe1}) is simplified to
\begin{equation}
\frac{\partial a_{d}(t)}{\partial t}=-\frac{i}{\hbar}F_{dc}e^{it(\omega
_{dc}-\omega)}-\frac{i}{\hbar}F_{dc}e^{it(\omega_{dc}+\omega)}] \label{coe2}%
\end{equation}
The solution of Eq.(\ref{coe2}) is simply a time integral:%
\begin{equation}
a_{d}(t)=-\frac{F_{dc}e^{it(\omega_{dc}-\omega)}}{\hbar(\omega_{dc}%
-\omega-i\delta)}-\frac{F_{dc}e^{it(\omega_{dc}+\omega)}}{\hbar(\omega
_{dc}+\omega-i\delta)},\text{ \ \ }\delta\rightarrow0^{+} \label{adc}%
\end{equation}
The change in wave function $\chi_{c}$ due to the ac field is%
\begin{equation}
\chi_{c}^{\prime}{}^{(1)}(t)=\sum_{d(\neq c)}(1-n_{d})a_{d}(t)\chi
_{d}e^{-itE_{d}/\hbar} \label{tic}%
\end{equation}
where $n_{d}$ is the occupation probability of single-electron state $\chi
_{d}$. The ``un-occupation" factor $(1-n_{d})$ is implicitly assumed when we
simplify Eq.(\ref{coe1}) to Eq.(\ref{coe2}): initially state $\chi_{d}$ must
be empty. In an intrinsic semiconductor at zero temperature, the valence band
is fully occupied and conduction band is empty. Eq.(\ref{tic}) indicates that
only conduction states can couple with the valence states. This implies that
the dc conductivity of an intrinsic semiconductor at zero temperature is zero.

With the help of Eq.(\ref{tp1}), Eq.(\ref{adc}) and Eq.(\ref{tic}),
Eq.(\ref{acc}) becomes%
\[
\mathbf{j}^{e}(\mathbf{r}^{\prime},t)=\frac{i\hbar e}{2m\Omega_{\mathbf{r}%
^{\prime}}}\int_{\Omega_{\mathbf{r}^{\prime}}}d\mathbf{s}\sum_{l_{1}%
l_{2}\cdots l_{N_{e}}}W_{l_{1}l_{2}\cdots l_{N_{e}}}\sum_{\alpha=1}^{N_{e}%
}\sum_{d(\neq l_{\alpha})}(1-n_{d})\frac{e}{2\hbar}\mathbf{E}_{0}\cdot
\]%
\begin{equation}
\{\langle\chi_{d}|\mathbf{r}|\chi_{l_{\alpha}}\rangle^{\ast}[\frac
{e^{it\omega}}{(\omega_{dl_{\alpha}}-\omega)}+\frac{e^{-it\omega}}%
{(\omega_{dl_{\alpha}}+\omega)}](\chi_{l_{\alpha}}\nabla_{\mathbf{s}}\chi
_{d}^{\ast}-\chi_{d}^{\ast}\nabla_{\mathbf{s}}\chi_{l_{\alpha}}) \label{cvec}%
\end{equation}%
\[
+\langle\chi_{d}|\mathbf{r}|\chi_{l_{\alpha}}\rangle\lbrack\frac{e^{-it\omega
}}{(\omega_{dl_{\alpha}}-\omega)}+\frac{e^{it\omega}}{(\omega_{dl_{\alpha}%
}+\omega)}](\chi_{d}\nabla_{\mathbf{s}}\chi_{l_{\alpha}}^{\ast}-\chi
_{l_{\alpha}}^{\ast}\nabla_{\mathbf{s}}\chi_{d})\}
\]
Separate out $\cos\omega t$ terms in Eq.(\ref{cvec}); they are in phase with
the external field. The real part of conductivity then reads:%
\begin{equation}
\sigma_{\mu\nu}^{(1)}(\omega)=\frac{e^{2}\hbar}{2m\Omega}\sum_{l_{1}%
l_{2}\cdots l_{N_{e}}}W_{l_{1}l_{2}\cdots l_{N_{e}}}\sum_{\alpha=1}^{N_{e}%
}\sum_{d(\neq l_{\alpha})}(1-n_{d})(\frac{1}{E_{l_{\alpha}}-E_{d}+\hbar\omega
}+\frac{1}{E_{l_{\alpha}}-E_{d}-\hbar\omega}) \label{cr}%
\end{equation}%
\[
\operatorname{Im}[\langle\chi_{d}|x_{\nu}|\chi_{l_{\alpha}}\rangle\int
_{\Omega}d^{3}x(\chi_{d}\frac{\partial\chi_{l_{\alpha}}^{\ast}}{\partial
x_{\mu}}-\chi_{l_{\alpha}}^{\ast}\frac{\partial\chi_{d}}{\partial x_{\mu}})]
\]
When $\omega=0$ Eq.(\ref{cr}) is reduces to Eq.(\ref{stc}) except for the
factor $(1-n_{d})$, as it should be.

To compare with the result from Boltzmann equation, let us consider a
crystalline metal. The semi-classical current density is given by\cite{ash}%
\begin{equation}
j_{\mu}^{e}=e\int\frac{2d\mathbf{k}}{(2\pi)^{3}}v_{\mu}(\mathbf{k}%
)g(\mathbf{k}) \label{cu}%
\end{equation}
where the integral is over the first Brillouin zone. $g(\mathbf{k})$ is the
non-equilibrium distribution function under external field, in relaxation time
approximation\cite{ash}%
\begin{equation}
g(\mathbf{k})=f(\mathbf{k})+\frac{\partial f}{\partial E}e\mathbf{E}%
\cdot\mathbf{v}\tau\lbrack E(\mathbf{k})] \label{bdis}%
\end{equation}
where $\tau\lbrack E(\mathbf{k})]$ is the energy-dependent relaxation time,
$f(\mathbf{k})$ is the Fermi distribution function $f[E(\mathbf{k})]$. The
conductivity is read out\cite{ash} from Eqs.(\ref{cu}) and (\ref{bdis}):%
\begin{equation}
\sigma_{\mu\nu}=e^{2}\int\frac{2d\mathbf{k}}{(2\pi)^{3}}\frac{\partial
f}{\partial E}\tau\lbrack E(\mathbf{k})]v_{\mu}(\mathbf{k})v_{\nu}%
(\mathbf{k}),\text{ \ \ }\mu,\nu=x,y,z \label{sej}%
\end{equation}
Using Eq.(\ref{uid}) and the definition of velocity operator $v_{\mu}%
=\frac{-i\hbar}{m}\frac{\partial}{\partial x_{\mu}}$, Eq.(\ref{cr}) is changed into%

\[
\sigma_{\mu\nu}^{(1)}(\omega)=\frac{e^{2}}{\Omega}\sum_{l_{1}l_{2}\cdots
l_{N_{e}}}W_{l_{1}l_{2}\cdots l_{N_{e}}}\sum_{\alpha=1}^{N_{e}}\sum_{d(\neq
l_{\alpha})}(1-n_{d})(\frac{1}{E_{l_{\alpha}}-E_{d}+\hbar\omega}+\frac
{1}{E_{l_{\alpha}}-E_{d}-\hbar\omega})\frac{\hbar}{E_{l_{\alpha}}-E_{d}}%
\]%
\begin{equation}
\operatorname{Im}\{[\int_{\Omega}d^{3}x\chi_{d}^{\ast}v_{\nu}\chi_{l_{\alpha}%
}][\frac{1}{2}\int_{\Omega}d^{3}x(\chi_{l_{\alpha}}^{\ast}v_{\mu}\chi_{d}%
-\chi_{d}v_{\mu}\chi_{l_{\alpha}}^{\ast})]\} \label{BL}%
\end{equation}
If we interpret $\frac{\hbar}{E_{l_{\alpha}}-E_{d}}$ as the energy-dependent
relaxation time $\tau(E_{l_{\alpha}})$ caused by inelastic scattering of
phonons (in a given MD step, scattering is caused by deviation from crystal;
to reflect various vibration states and electron-phonon scattering, averaging
over many MD steps is necessary\cite{gali,abt}), $n_{l_{\alpha}}%
(1-n_{d})(\frac{1}{E_{l_{\alpha}}-E_{d}+\hbar\omega}+\frac{1}{E_{l_{\alpha}%
}-E_{d}-\hbar\omega})$ as $\frac{\partial f}{\partial E}$, sum over states as
integral over Brillouin zone: $\frac{1}{\Omega}\sum_{l_{1}l_{2}\cdots
l_{N_{e}}}W_{l_{1}l_{2}\cdots l_{N_{e}}}\sum_{\alpha=1}^{N_{e}}\sum_{d(\neq
l_{\alpha})}\rightarrow\int\frac{2d\mathbf{k}}{(2\pi)^{3}}$ (only in crystal,
one can use $\mathbf{k}$-points in the reciprocal space to characterize
states), where $n_{l_{\alpha}}$ is removed from $W$. At $\omega=0,$ as
expected, Eq.(\ref{BL}) is reduced to the semi-classical result (\ref{sej}).

The $\sin\omega t$ terms terms in Eq.(\ref{cvec}) lag 90$^{0}$ behind the
phase of external field. The contribution to the imaginary part of
conductivity is%
\[
\sigma_{\alpha\beta}^{(2)}(\omega)=\frac{e^{2}\hbar}{2m\Omega}\sum_{l_{1}%
l_{2}\cdots l_{N_{e}}}W_{l_{1}l_{2}\cdots l_{N_{e}}}\frac{1}{N_{e}}%
\sum_{\alpha=1}^{N_{e}}\sum_{d(\neq l_{\alpha})}(1-n_{d})[\frac{1}%
{E_{d}-E_{l_{\alpha}}-\hbar\omega}-\frac{1}{E_{d}-E_{l_{\alpha}}+\hbar\omega
}]
\]%
\begin{equation}
\operatorname{Re}[\langle\chi_{d}|x_{\beta}|\chi_{l_{\alpha}}\rangle
\int_{\Omega}d^{3}x(\chi_{d}\frac{\partial\chi_{l_{\alpha}}^{\ast}}{\partial
x_{\alpha}}-\chi_{l_{\alpha}}^{\ast}\frac{\partial\chi_{d}}{\partial
x_{\alpha}})] \label{ime}%
\end{equation}
it is interesting to notice that $\sigma_{\alpha\beta}^{(2)}(0)=0$.

To first order in the field, the second term in Eq.(\ref{mdliu}) is%

\begin{equation}
-\frac{e^{2}}{m}\mathbf{A}(\mathbf{r},t)n_{\{\mathbf{W}_{1}\cdots
\mathbf{W}_{\mathcal{N}}\}}^{e}(\mathbf{r},t)=\frac{e^{2}\mathbf{E}_{0}%
}{m\omega}n_{\{\mathbf{W}_{1}\cdots\mathbf{W}_{\mathcal{N}}\}}^{e}%
(\mathbf{r})\sin\omega t \label{2si}%
\end{equation}
From Eq.(\ref{force}), to 1st order of field, the 3rd term in Eq.(\ref{mdliu}) is%

\begin{equation}
\sum_{\alpha}\frac{q_{\alpha}^{2}\mathbf{E}_{0}}{M_{\alpha}\omega}\sin\omega
t\delta(\mathbf{r}-\mathbf{W}_{\alpha}) \label{nuc}%
\end{equation}
Using Eq.(\ref{sap}), the imaginary part of conductivity of the electrons +
nuclei is%
\begin{equation}
\sigma_{\alpha\beta}^{I}(\omega)=\sigma_{\alpha\beta}^{(2)}(\omega
)+\frac{e^{2}n^{e}}{m\omega}+\sum_{p}\frac{q_{p}^{2}n_{p}^{N}}{M_{p}\omega}
\label{imc}%
\end{equation}
$M_{p}$, $q_{p}$ and $n_{p}^{N}$ are mass, effective charge and the number
density of the $p^{\text{th}}$ species of nuclei. The last two terms are
contributions from free charges\cite{jac}.

We cannot use Eq.(\ref{adc}) in two situations: (1) degenerate states with low
frequency field, $\omega_{dc}=0$ and $\omega\rightarrow0$; (2) external field
and two groups of levels in resonance: $\omega_{dc}-\omega=0$ or $\omega
_{dc}+\omega=0$. In these situations, Eq.(\ref{coe2}) leads to $a_{d}%
(t)\thicksim t$.

\subsubsection{Interaction of degenerate states with a very low frequency
external field}

For a group of $M$ degenerate states ($\chi_{d_{\sigma}},$ $\sigma
=1,2,\cdots,M)$, the mutual coupling is much stronger than the coupling
between one member and the states with different energy. The general evolution
equation%
\begin{equation}
i\hbar\frac{da_{j}}{dt}e^{-iE_{j}t/\hbar}=\sum_{k}a_{k}V_{jk}e^{-iE_{k}%
t/\hbar},\text{ \ \ }V_{jk}=\int d\mathbf{r}\chi_{j}^{\ast}H_{fm}(t)\chi_{k}
\label{ebn}%
\end{equation}
is simplified to%
\begin{equation}
i\hbar\frac{da_{j}(t)}{dt}=\sum_{k}a_{k}(t)G_{jk}\cos\omega t,\text{ \ }%
G_{jk}=\int d\mathbf{r}\chi_{j}^{\ast}(-e\mathbf{E}_{0}\cdot\mathbf{r}%
)\chi_{k}\text{\ \ \ }j,k=d_{1},d_{2},\cdots,d_{M} \label{edg}%
\end{equation}
Introduce new variable\ $s=\sin\omega t$, Eq.(\ref{edg}) becomes%
\begin{equation}
i\hbar\omega\frac{da_{j}(s)}{ds}=\sum_{k}a_{k}(s)G_{jk}, \label{rdg}%
\end{equation}
Notice $G_{jk}$ does not depend on time, taking Fourier transform%
\begin{equation}
a_{j}(s)=\int dpa_{pj}e^{-ips},\text{ \ \ \ \ }j=d_{1},d_{2},\cdots,d_{M}
\label{fta}%
\end{equation}
in both sides of Eq.(\ref{rdg}) leads to%
\begin{equation}
\sum_{k}a_{pj}(G_{jk}-\hbar\omega p\delta_{jk})=0 \label{scut}%
\end{equation}
Because ($G_{jk}$) is Hermitian, its eigenvalues $p_{1},p_{2},\cdots,p_{M}$
are real and the matrix ($a_{p_{\mu}d_{\alpha}}$) is unitary. The $M$ special
solutions of Eq.(\ref{edg}) are%
\begin{equation}
\chi_{p_{\mu}}^{\prime(0)}(t)=e^{-ip_{\mu}\sin\omega t}\sum_{\alpha=1}%
^{M}a_{p_{\mu}d_{\alpha}}\chi_{d_{\alpha}},\text{ \ \ }\mu=1,2,\cdots,M
\label{jie}%
\end{equation}
The matrix elements of perturbation $(-e\mathbf{E}_{0}\cdot\mathbf{r})$
relative to the new zero order wave functions are diagonal:%
\begin{equation}
\int dr\chi_{p_{\mu}}^{\prime(0)\ast}(-e\mathbf{E}_{0}\cdot\mathbf{r}%
)\chi_{p_{\nu}}^{\prime(0)}=G_{p_{\mu}p_{\nu}}\delta_{\mu\nu},\text{ \ }%
\mu,\nu=1,2,\cdots,M \label{dign}%
\end{equation}
Because matrix ($a_{p_{\mu}d_{\alpha}}$) is unitary, one has%
\begin{equation}
\sum_{\mu=1}^{M}[\chi_{p_{\mu}}^{\prime(0)}(t)\nabla\chi_{p_{\mu}}%
^{\prime(0)\ast}(t)-\chi_{p_{\mu}}^{\prime(0)\ast}(t)\nabla\chi_{p_{\mu}%
}^{\prime(0)}(t)]-\sum_{\alpha=1}^{M}[\chi_{d_{\alpha}}\nabla\chi_{d_{\alpha}%
}^{\ast}-\chi_{d_{\alpha}}^{\ast}\nabla\chi_{d_{\alpha}}]=0 \label{0dg}%
\end{equation}
From Eq.(\ref{acc}), the zeroth order correction to degenerate states does not
contribute to conductivity.

Unlike the KGF, where dc conductivity is obtained either by extrapolating from
optical conductivity\cite{gali} or by writing a separate code for zero
frequency\cite{md,Over,abt}, the present ac expression includes dc expression
in the obvious way. One may notice when $\omega=0,$ $\mathbf{j}_{s}^{e}=0$ and
$\mathbf{j}_{s2}^{e}=0$ (cf. Appendix A). Excepting the factor ($1-n$) which
would not appear from single-electron stationary perturbation theory, when
$\omega\rightarrow0,$ $\mathbf{j}_{c}^{e}$ is reduced into the second sum of
Eq.(\ref{uds}), $(\mathbf{j}_{c2}^{e}+\mathbf{j}_{0}^{e})$ is reduced into the
third sum of Eq.(\ref{uds}) (cf. Appendix A).

In Appendix A, we show that the contributions to current density or
conductivity from the degenerate states is finite. In the single-electron
states, the number of degenerate states in each degenerate manifold is much
smaller than the total number of non-degenerate states. The conductivity from
the coupling among degenerate states can therefore be neglected. One can use
Eqs.(\ref{cr}) and (\ref{ime}), in which only counts the coupling among
non-degenerate states and the coupling between degenerate states and
non-degenerate states.

\subsubsection{Resonance with external field}

Suppose in ($\chi_{l_{1}},\chi_{l_{2}},\cdots,\chi_{l_{N_{e}}}$) there is a
$M-$fold degenerate manifold ($\chi_{m_{1}},\chi_{m_{2}},\cdots,\chi_{m_{M}}$)
and a $M^{\prime}-$fold degenerate manifold ($\chi_{n_{1}},\chi_{n_{2}}%
,\cdots,\chi_{n_{M^{\prime}}}$) which are nearly in resonance with a finite
frequency $\omega$ external field: $E_{m}^{(0)}-E_{n}^{(0)}=\hbar
(\omega+\epsilon),$ \ \ $\epsilon<<\omega$ or $\epsilon=0$. The coupling with
other non-resonant states can be neglected. If we only consider the
interaction with the smallest oscillating frequency,%
\begin{equation}
V_{m_{j}n_{k}}(t)\thickapprox F_{m_{j}n_{k}}e^{it(\omega_{m_{j}n_{k}}-\omega
)}=F_{m_{j}n_{k}}e^{it\epsilon},\text{ \ }F_{m_{j}n_{k}}=\frac{1}{2}\int
d\mathbf{r}\chi_{m_{j}}^{\ast}(-e\mathbf{E}_{0}\cdot\mathbf{r})\chi_{n_{k}}
\label{bgh}%
\end{equation}
the general evolution equation%
\begin{equation}
i\hbar\frac{da_{m_{j}}}{dt}=\sum_{k=1}^{M^{\prime}}V_{m_{j}n_{k}}(t)a_{n_{k}%
},\text{\ \ }j=1,2,\cdots,M;\text{ \ \ }k=1,2,\cdots,M^{\prime} \label{mj}%
\end{equation}
is simplified to%
\begin{equation}
i\hbar\frac{da_{m_{j}}}{dt}=\sum_{k=1}^{M^{\prime}}F_{m_{j}n_{k}}%
e^{it\epsilon}a_{n_{k}} \label{son}%
\end{equation}
Similarly if one only takes the interaction with smallest oscillating
frequency%
\begin{equation}
V_{n_{k}m_{j}}(t)\thickapprox F_{m_{j}n_{k}}^{\ast}e^{it(\omega_{n_{k}m_{j}%
}+\omega)}=F_{m_{j}n_{k}}^{\ast}e^{-it\epsilon} \label{sina}%
\end{equation}
the general evolution equation%
\begin{equation}
i\hbar\frac{da_{n_{k}}}{dt}=\sum_{j=1}^{M}V_{n_{k}m_{j}}(t)a_{m_{j}}
\label{ge}%
\end{equation}
is simplified to%
\begin{equation}
i\hbar\frac{da_{n_{k}}}{dt}=\sum_{j=1}^{M}F_{m_{j}n_{k}}^{\ast}e^{-it\epsilon
}a_{m_{j}} \label{res}%
\end{equation}

Introduce new functions $b_{n_{k}}$: $a_{n_{k}}=b_{n_{k}}e^{-it\epsilon}$
($k=1,2,\cdots,M^{\prime}$), Eq.(\ref{son}) becomes%
\begin{equation}
i\hbar\frac{da_{m_{j}}}{dt}=\sum_{k=1}^{M^{\prime}}F_{m_{j}n_{k}}b_{n_{k}}
\label{1am}%
\end{equation}
Eq.(\ref{res}) becomes%
\begin{equation}
\overset{\cdot}{i\hbar b}_{n_{k}}=-\hbar\epsilon b_{n_{k}}+\sum_{j=1}%
^{M}F_{m_{j}n_{k}}^{\ast}a_{m_{j}} \label{1bk}%
\end{equation}
Eqs.(\ref{1am}) and (\ref{1bk}) can be rearranged into%
\begin{equation}
i\hbar\frac{d}{dt}V=RV \label{matr}%
\end{equation}
where%
\begin{equation}
R=\left(
\begin{array}
[c]{cc}%
0_{M\times M} & B_{M\times M^{\prime}}\\
\lbrack(B_{M\times M^{\prime}})^{\text{transpose}}]^{\ast} & 0_{M^{\prime
}\times M^{\prime}}-\hbar\epsilon I_{M^{\prime}\times M^{\prime}}%
\end{array}
\right)  \label{rm}%
\end{equation}
is a $(M+M^{\prime})\times(M+M^{\prime})$ matrix, $0$ is zero matrix, $I$ is
the unit matrix, elements of matrix $B$ are given by%
\begin{equation}
B_{jk}=F_{m_{j}n_{k}},\text{ \ \ }j=1,2,\cdots,M;\text{ \ \ }k=1,2,\cdots
,M^{\prime} \label{mb}%
\end{equation}
$V$ is a $(M+M^{\prime})-$column vector, its transpose is%
\begin{equation}
V^{\text{transpose}}=(a_{m_{1}},a_{m2},\cdots,a_{m_{M}};b_{n_{1}},b_{n_{2}%
},\cdots,b_{n_{M^{\prime}}}) \label{vt}%
\end{equation}

We are looking for special solutions of Eq.(\ref{matr}) in the form:%
\begin{equation}
a_{m_{j}}^{q}(t)=a_{m_{j}}^{q0}e^{it\alpha_{q}},\text{ \ }j=1,2,\cdots
,M;\text{ \ \ \ }b_{n_{k}}^{q}(t)=b_{n_{k}}^{q0}e^{it\alpha_{q}},\text{
\ }k=1,2,\cdots,M^{\prime} \label{sh}%
\end{equation}
The column vector $V_{0}^{q}$ with%
\begin{equation}
\left(  V_{0}^{q}\right)  ^{\text{transpose}}=(a_{m_{1}}^{q0},a_{m2}%
^{q0},\cdots,a_{m_{M}}^{q0};b_{n_{1}}^{q0},b_{n_{2}}^{q0},\cdots
,b_{n_{M^{\prime}}}^{q0}) \label{0vt}%
\end{equation}
is the eigenvector of $R$ belonging to eigenvalue $\hbar\alpha_{q}$. Then
$(M+M^{\prime})$ special solutions of the time-dependent single-electron
Schrodinger equation are%
\begin{equation}
\chi_{q}^{\prime(0)}(t)=\sum_{j=1}^{M}a_{m_{j}}^{q0}e^{i\alpha_{q}t}%
\chi_{m_{j}}e^{-iE_{m}t/\hbar}+\sum_{k=1}^{M^{\prime}}b_{n_{k}}^{q0}%
e^{i(\alpha_{q}-\epsilon)t}\chi_{n_{k}}e^{-iE_{n}t/\hbar},\text{
\ }q=1,2,\cdots,M+M^{\prime} \label{SS}%
\end{equation}
and the general solution can be obtained from linear combinations. Because $R$
is Hermitian, its eigenvalues $(\hbar\alpha_{q},q=1,2,\cdots,M+M^{\prime})$
are real, matrix $C=(a_{m_{1}}^{q0},a_{m2}^{q0},\cdots,a_{m_{M}}^{q0}%
;b_{n_{1}}^{q0},b_{n_{2}}^{q0},\cdots,b_{n_{M^{\prime}}}^{q0})$ is unitary
($q$ is index of row).

If we use Eq.(\ref{SS}) and notice that $C$ is unitary, one has%
\begin{equation}
\sum_{q=1}^{M+M^{\prime}}(\chi_{q}^{\prime(0)}\nabla\chi_{q}^{\prime(0)\ast
}-\chi_{q}^{\prime(0)\ast}\nabla\chi_{q}^{\prime(0)})=\sum_{j=1}^{M}%
(\chi_{m_{j}}\nabla\chi_{m_{j_{1}}}^{\ast}-\chi_{m_{j_{1}}}^{\ast}\nabla
\chi_{m_{j}})+\sum_{k=1}^{M^{\prime}}(\chi_{n_{k}}\nabla\chi_{n_{k_{1}}}%
^{\ast}-\chi_{n_{k_{1}}}^{\ast}\nabla\chi_{n_{k}}) \label{zr}%
\end{equation}
From Eq.(\ref{sux}), the contribution to current from two groups of states in
resonance with an external field is zero. The artificial poles in the case of
resonance in Eq.(\ref{adc}) are removed. For a mechanical oscillator, if we
input energy in a resonant way and do not take out energy, the amplitude of
the oscillator will increase indefinitely. The situation for two groups of
resonant levels is different; the system absorbs the external field while
stimulated emission also occurs. The material and field are in
absorption-emission equilibrium, so that no singularity occurs. The two
$\delta$ functions (they originate from first order correction of wave
function) in Greenwood formula come from the long time limit, and are not
caused by resonance. In Appendix B, we show that the contribution from
resonant states is finite.

\subsection{Comparison with Greenwood formula}

Both the present work and the Greenwood derivation require the use of
perturbation theory:%
\begin{equation}
a_{d}(t)=-\frac{i}{\hbar}F_{dc}\int_{t_{1}}^{t_{2}}dt^{\prime}[e^{it^{\prime
}(\omega_{dc}-\omega)}+e^{it^{\prime}(\omega_{dc}+\omega)}]<<1 \label{pertc}%
\end{equation}
This means that the interaction time $\tau=(t_{2}-t_{1})$ cannot be too long:%
\begin{equation}
\tau=(t_{2}-t_{1})<<\frac{\hbar}{F} \label{ptc}%
\end{equation}
The Greenwood derivation also requires that the transition probability
\textit{per unit time} be defined:%
\begin{equation}
\lim_{\tau\rightarrow\infty}\frac{\sin^{2}\frac{T(\omega_{dc}-\omega)}{2}%
}{\tau(\frac{\omega_{dc}-\omega}{2})^{2}}=\pi\delta(\frac{\omega_{dc}-\omega
}{2})\text{ and \ \ }\lim_{\tau\rightarrow\infty}\frac{\sin^{2}\frac
{T(\omega_{dc}+\omega)}{2}}{\tau(\frac{\omega_{dc}+\omega}{2})^{2}}=\pi
\delta(\frac{\omega_{dc}+\omega}{2}) \label{tad}%
\end{equation}
That is, the interaction time $\tau$ should be long enough%
\begin{equation}
\tau>>\frac{2}{\omega_{dc}-\omega}\text{ and }\tau>>\frac{2}{\omega
_{dc}+\omega} \label{jc}%
\end{equation}
to allow the two limits in Eq.(\ref{tad}) to be taken. The law of conservation
energy (of field + matter) can be verified by means of two measurements only
to an \textit{accuracy} of the order of $\hbar/\Delta t,$ where $\Delta t$ is
the time interval between the measurements\cite{ll}, i.e. the interaction time
$\tau$ between field and matter. Since in present work we do not need
probability per unit time (i.e. long time limit), the energy conservation
delta function will not appear.

For a large system with continuous energy spectrum, Eq.(\ref{jc}) contradicts
Eq.(\ref{ptc}) for close levels when $\omega\rightarrow0$. Since a transition
with small $\omega_{dc}$ makes a large contribution to conductivity, the dc
conductivity obtained from the KGF is problematic. The derivation in this work
does not need transition probability, therefore does not need condition
(\ref{jc}), and is self-consistent. Numerically Eq.(\ref{cr}) and Eq.
(\ref{imc}) get rid of the delta function, and do not require an artificial broadening.

\section{Role of many-electron statistics}

In this section, method (2), many-body perturbation theory, is used to compute
the conductivity. We take a static field as example and apply to concrete examples.

\subsection{Intrinsic semiconductor}

\subsubsection{K-electron excited state}

Label the states in valence band from low energy to high energy as $v_{N_{e}%
},\cdots,v_{2},v_{1}$, the states in the conduction band from low energy to
high energy as $c_{1},c_{2},\cdots,c_{N_{e}}$. For an intrinsic semiconductor
at $T=$0K, the valence band is full, and the number of states is the number of
electrons. The system is in its ground state%
\begin{equation}
\Lambda_{0}=\frac{1}{\sqrt{N_{e}!}}\sum_{P}\delta_{P}P_{r_{1}s_{z1}%
,r_{2}s_{z2}\cdots r_{N_{e}}s_{zN_{e}}}v_{1}(r_{1}s_{z1})v_{2}(r_{2}%
s_{z2})\cdots v_{N_{e}}(r_{N_{e}}s_{zN_{e}}) \label{gs}%
\end{equation}

At $T>0$, various excited states appear. First consider one electron pumped
into the conduction band from\ the valence band. A one-electron excited state
$\Lambda_{v_{j}c_{k}}$ is constructed from $\Lambda_{0}$ by replacing $v_{j}$
with $c_{k}$. There are $N_{e}$ manners in choosing $v_{j}$. There are $N_{e}$
manners in choosing $c_{k}$. The single-electron energy spectrum is very
dense, there are many combinations of the choices of initial valence state and
the final conduction state, the degeneracy among 1-electron excited states is high.

Since any observable, including current, are bilinear forms of a $N_{e}%
$-electron wave function, the order of $N_{e}$ states in the Slater
determinant $\Lambda_{v_{j}c_{k}}$ does not matter provided that we maintain
the same order in $\Lambda_{v_{j}c_{k}}^{\ast}$. The probability of state
$\Lambda_{v_{j}c_{k}}$ relative to the ground state $\Lambda_{0}$ is%
\begin{equation}
U_{v_{j}c_{k}}=\exp\{-[(E_{v_{1}}-E_{v_{j}})+E_{g}+(E_{c_{k}}-E_{c_{1}%
})\}/(k_{B}T)\} \label{r1}%
\end{equation}
where $E_{g}$ is the band gap.

If two electrons are pumped from valence band to conduction band, a 2-electron
excited state $\Lambda_{v_{j_{1}}v_{j_{2}}c_{p_{1}}c_{p_{2}}}$ is obtained
from $\Lambda_{0}$ by replacing ($v_{j_{1}},v_{j_{2}}$) with ($c_{p_{1}%
},c_{p_{2}}$). The relative probability of state $\Lambda_{v_{j_{1}}v_{j_{2}%
}c_{p_{1}}c_{p_{2}}}$ to ground state $\Lambda_{0}$ is%
\begin{equation}
U_{v_{j_{1}}v_{j_{2}}c_{p_{1}}c_{p_{2}}}=\exp\{-[(E_{v_{1}}-E_{v_{j_{1}}%
})+(E_{v_{1}}-E_{v_{j_{2}}})+2E_{g}+(E_{c_{p_{1}}}-E_{c_{1}})+(E_{c_{p_{2}}%
}-E_{c_{1}})]/(k_{B}T)] \label{r2}%
\end{equation}

In $\Lambda_{0}$ if K electrons are excited to the conduction band from
valence band: state $v_{j_{1}}$ is replaced by state $c_{p_{1}}$, state
$v_{j_{2}}$ is replaced by $c_{p_{2}}$, $\cdots,$ state $v_{j_{K}}$ is
replaced by state $c_{p_{K}}$, a K-electron excited state $\Lambda_{v_{j_{1}%
}v_{j_{2}}\cdots v_{j_{K}};c_{p_{1}}c_{p_{2}}\cdots c_{p_{K}}}$ is obtained.
The probability of $\Lambda_{v_{j_{1}}v_{j_{2}}\cdots v_{j_{K}}c_{p_{1}%
}c_{p_{2}}\cdots c_{p_{K}}}$ relative to $\Lambda_{0}$ is%
\begin{equation}
U_{v_{j_{1}}v_{j_{2}}\cdots v_{j_{K}}c_{p_{1}}c_{p_{2}}\cdots c_{p_{K}}}%
=\exp\{-[\sum_{\alpha=1}^{K}(E_{v_{1}}-E_{v_{j_{^{\alpha}}}})+KE_{g}%
+\sum_{\alpha=1}^{K}(E_{c_{_{p_{\alpha}}}}-E_{c_{1}})]/(k_{B}T)\} \label{rK}%
\end{equation}

The absolute probability $W_{v_{j}c_{k}}$ of 1-electron excited state
$\Lambda_{v_{j}c_{k}}$ is%
\begin{equation}
W_{v_{j}c_{k}}=\frac{U_{v_{j}c_{k}}}{Z},\text{ \ \ }Z=1+\sum_{jk}U_{v_{j}%
c_{k}}+\cdots+\sum_{v_{j_{1}}v_{j_{2}}\cdots v_{j_{K}};c_{p_{1}}c_{p_{2}%
}\cdots c_{p_{K}}}U_{v_{j_{1}}v_{j_{2}}\cdots v_{j_{K}}c_{p_{1}}c_{p_{2}%
}\cdots c_{p_{K}}}+\cdots\label{a1}%
\end{equation}
It is easy to verify for low $T$ that $\exp(\frac{E_{V}-E_{F}}{k_{B}T})<<1$
and $\exp(\frac{E_{c}-E_{F}}{k_{B}T})>>1$ (they are satisfied even in several
thousand K), so that one has%
\begin{equation}
W_{v_{j}c_{k}}=[1-f(E_{v_{j}})]f(E_{c_{k}}) \label{stc1}%
\end{equation}
where%
\begin{equation}
f(E_{v})=[\exp(\frac{E_{V}-E_{F}}{k_{B}T})+1]^{-1}\text{ and }f(E_{c}%
)=[\exp(\frac{E_{c}-E_{F}}{k_{B}T})+1]^{-1} \label{fer}%
\end{equation}
are Fermi distribution functions of valence states and conduction states.

The absolute probability $W_{v_{j_{1}}v_{j_{2}}c_{p_{1}}c_{p_{2}}}$ of
2-electron excited state $\Lambda_{v_{j_{1}}v_{j_{2}}c_{p_{1}}c_{p_{2}}}$ can
be obtained similarly%
\begin{equation}
W_{v_{j_{1}}v_{j_{2}}c_{p_{1}}c_{p_{2}}}=U_{v_{j_{1}}v_{j_{2}}c_{p_{1}%
}c_{p_{2}}}/Z=[1-f(E_{v_{j_{1}}})][1-f(E_{v_{j_{2}}})]f(E_{c_{p_{1}}%
})f(E_{c_{p_{2}}}) \label{2w}%
\end{equation}
The absolute probability $W_{v_{j_{1}}v_{j_{2}}\cdots v_{j_{K}}c_{p_{1}%
}c_{p_{2}}\cdots c_{p_{K}}}$ of K-electron excited state $\Lambda_{v_{j_{1}%
}v_{j_{2}}\cdots v_{j_{K}};c_{p_{1}}c_{p_{2}}\cdots c_{p_{K}}}$ is%
\begin{equation}
W_{v_{j_{1}}v_{j_{2}}\cdots v_{j_{K}}c_{p_{1}}c_{p_{2}}\cdots c_{p_{K}}%
}=U_{v_{j_{1}}v_{j_{2}}\cdots v_{j_{K}}c_{p_{1}}c_{p_{2}}\cdots c_{p_{K}}}/Z=%
{\displaystyle\prod\limits_{\alpha=1}^{K}}
[1-f(E_{v_{j_{\alpha}}})]f(E_{c_{p_{\alpha}}}) \label{Kp}%
\end{equation}

\subsubsection{Zero dc conductivity at T=0K}

Because the interaction with a static field%
\begin{equation}
H_{int}=-\sum_{m=1}^{N_{e}}e\mathbf{E}\cdot\mathbf{r}_{m} \label{in1}%
\end{equation}
is a single-particle operator (separable for coordinate\ of each particle),
the ground state only couples with 1-electron excited states%
\begin{equation}
\langle\Lambda_{v_{j}c_{k}}|-\sum_{m=1}^{N_{e}}e\mathbf{E}\cdot\mathbf{r}%
_{m}|\Lambda_{0}\rangle=\int d\mathbf{r}_{1}c_{k}^{\ast}(\mathbf{r}%
_{1})(-e\mathbf{E}\cdot\mathbf{r}_{1})v_{j}(\mathbf{r}_{1}) \label{01}%
\end{equation}
The change in ground state $\Lambda_{0}$ by external field only includes
1-electron excited states
\begin{equation}
\Lambda_{0}^{\prime(1)}=\sum_{jk}\frac{\langle\Lambda_{v_{j}c_{k}}|-\sum
_{m=1}^{N_{e}}e\mathbf{E}\cdot\mathbf{r}_{m}|\Lambda_{0}\rangle}%
{E_{0}-E_{v_{j}c_{k}}}\Lambda_{v_{j}c_{k}}=\sum_{jk}\frac{\langle
c_{k}(1)|e\mathbf{E}\cdot\mathbf{r}_{1}|v_{j}(1)\rangle}{(E_{C_{k}}-E_{c_{1}%
})+(E_{v_{1}}-E_{v_{j}})+E_{g}}\Lambda_{v_{j}c_{k}} \label{1w0}%
\end{equation}

If we take Eq.(\ref{mdliu}) and effect the multiple integral, the current
density is%
\begin{equation}
\mathbf{j}^{e}(\mathbf{r})=\frac{ie\hbar N_{e}}{2m\Omega}\{\sum_{jk}%
\frac{\langle c_{k}(1)|e\mathbf{E}\cdot\mathbf{r}_{1}|v_{j}(1)\rangle^{\ast}%
}{(E_{C_{k}}-E_{c_{1}})+(E_{v_{1}}-E_{v_{j}})+E_{g}}\int_{\Omega_{\mathbf{r}}%
}d\mathbf{s}(v_{j}(\mathbf{s})\nabla_{\mathbf{s}}c_{k}^{\ast}(\mathbf{s}%
)-c_{k}^{\ast}(\mathbf{s})\nabla_{\mathbf{s}}v_{j}(\mathbf{s})) \label{mic}%
\end{equation}%
\[
+\sum_{jk}\frac{\langle c_{k}(1)|e\mathbf{E}\cdot\mathbf{r}_{1}|v_{j}%
(1)\rangle}{(E_{C_{k}}-E_{c_{1}})+(E_{v_{1}}-E_{v_{j}})+E_{g}}\int
_{\Omega_{\mathbf{r}}}d\mathbf{s}(c_{k}(\mathbf{s})\nabla_{\mathbf{s}}%
v_{j}^{\ast}(\mathbf{s})-v_{j}^{\ast}(\mathbf{s})\nabla_{\mathbf{s}}%
c_{k}(\mathbf{s}))\}
\]
Using Eq.(\ref{cdu}), one can read off conductivity. Because the external
field is much weaker than the atomic field, the numerator is much smaller than
the energy gap $E_{g}$ (this will become more obvious in next section), the
change $\Lambda_{0}^{\prime(1)}$ in wave function $\Lambda_{0}$ can be
neglected, and the dc conductivity is negligible at T=0 in an intrinsic
semiconductor. The coupling between 0-electron to 1-electron excited states
can be viewed as a cross band transition, its probability is not exactly zero,
but is extremely small. One may neglect the existence of conduction band:
electron cannot be accelerated when valence band is full. To accelerate an
electron in ground state, one has to go from valence band to conduction band.
The probability is negligible for an external field which is much weaker than
atomic field.

\subsubsection{Conduction from one-electron excited states}

Because $H_{int}$ is single-particle operator, a 1-electron excited state
could couple with ground state, 1-electron excited states and 2-electron
excited states. \bigskip The energy difference between a 1-electron excited
state and a 2-electron excited state is at least energy gap $E_{g}$. The
contribution to current density from this coupling is small. So does the
coupling between a 1-electron excited state and $\Lambda_{0}$.

Since $H_{int}$ is a single-particle operator, there are only two types of
coupling between two different 1-electron excited states: $\Lambda_{v_{j}%
c_{k}}\leftrightarrow\Lambda_{v_{j}c_{k^{\prime}}}$ \ or \ $\Lambda
_{v_{j}c_{k}}\leftrightarrow\Lambda_{v_{j^{\prime}}c_{k}}$. The energy
difference between such pairs of 1-electron excited states can be small if
states $c_{k^{\prime}}$ and $c_{k}$ ($v_{j^{\prime}}$ and $v_{j}$) are
properly chosen. Their contribution will be much larger than the
coupling\ between a K-electron excited state and a ($K\pm1$)-electron excited state.

The change in $\Lambda_{v_{j}c_{k}}$ caused by a static field is%
\begin{equation}
\Lambda_{v_{j}c_{k}}^{^{\prime}(1)}=\sum_{k^{\prime}}\frac{\langle
c_{k^{\prime}}(1)|e\mathbf{E}\cdot\mathbf{r}_{1}|c_{k}(1)\rangle
}{E_{c_{k^{\prime}}}-E_{c_{k}}}\Lambda_{v_{j}c_{k^{\prime}}}+\sum_{j^{\prime}%
}\frac{\langle v_{j^{\prime}}(1)|e\mathbf{E}\cdot\mathbf{r}_{1}|v_{j}%
(1)\rangle}{E_{v_{j^{\prime}}}-E_{v_{j}}}\Lambda_{v_{j^{\prime}}c_{k}}
\label{c11}%
\end{equation}
By appealing to Eqs. (\ref{urc}), (\ref{cdu}) and (\ref{uid}), the expression
of conductivity in momentum representation is%
\[
\sigma_{\alpha\beta}=\frac{e^{2}\hbar^{3}N_{e}}{m^{2}\Omega_{\mathbf{r}%
^{\prime}}}\sum_{jk}W_{v_{j}c_{k}}\operatorname{Im}\{\sum_{k^{\prime}}%
\frac{\langle c_{k^{\prime}}(1)|\frac{\partial}{\partial x_{1\beta}}%
|c_{k}(1)\rangle}{(E_{c_{k^{\prime}}}-E_{c_{k}})^{2}}\int_{\Omega
_{\mathbf{r}^{\prime}}}d\mathbf{r}(c_{k^{\prime}}(\mathbf{r})\frac{\partial
c_{k}^{\ast}(\mathbf{r})}{\partial x_{\alpha}}-c_{k}^{\ast}(\mathbf{r}%
)\frac{\partial c_{k^{\prime}}(\mathbf{r})}{\partial x_{\alpha}})
\]%
\begin{equation}
+\sum_{j^{\prime}}\frac{\langle v_{j^{\prime}}(1)|\frac{\partial}{\partial
x_{1\beta}}|v_{j}(1)\rangle}{(E_{v_{j^{\prime}}}-E_{v_{j}})^{2}}\int
_{\Omega_{\mathbf{r}^{\prime}}}d\mathbf{r}(v_{j^{\prime}}(\mathbf{r}%
)\frac{\partial v_{j}^{\ast}(\mathbf{r})}{\partial x_{\alpha}}-v_{j}^{\ast
}(\mathbf{r})\frac{\partial v_{j^{\prime}}(\mathbf{r})}{\partial x_{\alpha}%
})\} \label{ong}%
\end{equation}
The accelerated hole in the valence band and the accelerated electron in
conduction band contribute most to the conduction, the coupling between
$K$-electron and ($K\pm1$)-electron excited states contributes much less. We
have proven that degenerate states act like non-degenerate states, cf.
Eq.(\ref{uds}). All the denominators in Eq.(\ref{ong}) are not zero. Except
delta functions, Eq.(\ref{ong}), the contribution from 1-electron excited
states, corresponds to the ordinary Greenwood formula.

In the standard procedure applying KGF\cite{Over}%
\begin{equation}
\sigma(T)=\int_{-\infty}^{\infty}dE\sigma(E)[-\frac{df}{dE}],\text{ \ }%
\sigma(E)=\frac{\pi\hbar^{2}}{\Omega m^{2}}\sum_{mn}|\langle n|p_{x}%
|m\rangle|^{2}\delta(E_{n}-E)\delta(E_{m}-E) \label{KGF}%
\end{equation}
one broadens delta function by a Gaussian%
\begin{equation}
\delta(E_{n}-E)\thickapprox\frac{\exp[-(E_{n}-E)^{2}/(2\Delta^{2})]}%
{\Delta\sqrt{2\pi}} \label{Gau}%
\end{equation}
Numerically, this procedure is equivalent to replace whole series about
$(E_{v_{j^{\prime}}}-E_{v_{j}})^{-2}$ in Eq.(\ref{ong}) with several large
terms, each of order of $\Delta^{-2}$. There are two relevant energy scales in
the problem: $k_{B}T$, and characteristic energy level splittings near $E_{F}%
$. The choices of $\Delta$ is thus somewhat arbitrary. On one hand $\Delta$
should be order of $k_{B}T$ to reflect thermal environment. However $k_{B}T$
is a too small choice of $\Delta$ for room temperature but may be too large
for a high temperature. On the other hand, $\Delta$ should be order of or
smaller than the eigenvalue splittings near E$_{F}$. This choice depends on
the size of a structural model and also depends on how many k--points one
wishes to use. Thus KGF depends on a fortunate choice of $\Delta$, or requires
some other extrapolation scheme to $\omega=0$. Eq.(\ref{ong}) or Eq.(\ref{cr})
does not suffer from this problem.

\subsubsection{Conduction from 2-electron excited states}

Although a 2-electron excited state may couple with a 1-electron excited state
or a 3-electron excited state, the energy differences are at least energy gap
$E_{g}$. Later, we only consider the matrix elements between two 2-electron
excited states. The 1$^{\text{st}}$ order correction to $\Lambda_{v_{j_{1}%
}v_{j_{2}}c_{p_{1}}c_{p_{2}}}$ is%
\[
\Lambda_{v_{j_{1}}v_{j_{2}}c_{p_{1}}c_{p_{2}}}^{\prime(1)}=\sum_{j_{1}%
^{\prime}}\frac{\langle v_{j_{1}^{\prime}}(1)|e\mathbf{E}\cdot\mathbf{r}%
_{1}|v_{j_{1}}(1)\rangle}{E_{v_{j_{1}^{\prime}}}-E_{v_{j_{1}}}}\Lambda
_{v_{j_{1}^{\prime}}v_{j_{2}}c_{p_{1}}c_{p_{2}}}+\sum_{j_{2}^{\prime}}%
\frac{\langle v_{j_{2}^{\prime}}(1)|e\mathbf{E}\cdot\mathbf{r}_{1}|v_{j_{2}%
}(1)\rangle}{E_{v_{j_{2}^{\prime}}}-E_{v_{j_{2}}}}\Lambda_{v_{j_{1}}%
v_{j_{2}^{\prime}}c_{p_{1}}c_{p_{2}}}%
\]%
\begin{equation}
+\sum_{p_{1}^{\prime}}\frac{\langle c_{p_{1}^{\prime}}(1)|e\mathbf{E}%
\cdot\mathbf{r}_{1}|c_{p_{1}}(1)\rangle}{E_{c_{p_{1}^{\prime}}}-E_{c_{p_{1}}}%
}\Lambda_{v_{j_{1}}v_{j_{2}}c_{p_{1}^{\prime}}c_{p_{2}}}+\sum_{p_{2}^{\prime}%
}\frac{\langle c_{p_{2}^{\prime}}(1)|e\mathbf{E}\cdot\mathbf{r}_{1}|c_{p_{2}%
}(1)\rangle}{E_{c_{p_{2}^{\prime}}}-E_{c_{p_{2}}}}\Lambda_{v_{j_{1}}v_{j_{2}%
}c_{p_{1}}c_{p_{2}^{\prime}}} \label{yc2}%
\end{equation}
where we keep the single-electron wave functions in each Slater determinant in
a fixed order. Substitute Eq.(\ref{yc2}) into Eq.(\ref{mdliu}) and effect the
multiple integral, the macroscopic current density is then:%
\[
\mathbf{j}(\mathbf{r}^{\prime})=\frac{ie\hbar N_{e}}{2m\Omega_{\mathbf{r}%
^{\prime}}}\sum_{j_{1},j_{2}(>j_{1})}\sum_{p_{1},p_{2}(>p_{1})}W_{v_{j_{1}%
}v_{j_{2}}c_{p_{1}}c_{p_{2}}}%
\]%
\[
\{\sum_{j_{1}^{\prime}(\neq j_{1})}\frac{\langle v_{j_{1}^{\prime}%
}(1)|e\mathbf{E}\cdot\mathbf{r}_{1}|v_{j_{1}}(1)\rangle^{\ast}}{E_{v_{j_{1}%
^{\prime}}}-E_{v_{j_{1}}}}\int_{\Omega_{\mathbf{r}^{\prime}}}d\mathbf{r}%
(v_{j_{1}}(\mathbf{r})\nabla_{\mathbf{r}}v_{j_{1}^{\prime}}^{\ast}%
(\mathbf{r})-v_{j_{1}^{\prime}}^{\ast}(\mathbf{r})\nabla_{\mathbf{r}}v_{j_{1}%
}(\mathbf{r}))
\]%
\[
+\sum_{j_{2}^{\prime}(\neq j_{2})}\frac{\langle v_{j_{2}^{\prime}%
}(1)|e\mathbf{E}\cdot\mathbf{r}_{1}|v_{j_{2}}(1)\rangle^{\ast}}{E_{v_{j_{2}%
^{\prime}}}-E_{v_{j_{2}}}}\int_{\Omega_{\mathbf{r}^{\prime}}}d\mathbf{r}%
(v_{j_{2}}(\mathbf{r})\nabla_{\mathbf{r}}v_{j_{2}^{\prime}}^{\ast}%
(\mathbf{r})-v_{j_{2}^{\prime}}^{\ast}(\mathbf{r})\nabla_{\mathbf{r}}v_{j_{2}%
}(\mathbf{r}))
\]%
\[
+\sum_{p_{1}^{\prime}(\neq p_{1})}\frac{\langle c_{p_{1}^{\prime}%
}(1)|e\mathbf{E}\cdot\mathbf{r}_{1}|c_{p_{1}}(1)\rangle^{\ast}}{E_{c_{p_{1}%
^{\prime}}}-E_{c_{p_{1}}}}\int_{\Omega_{\mathbf{r}^{\prime}}}d\mathbf{r}%
(c_{p_{1}}(\mathbf{r})\nabla_{\mathbf{r}}c_{p_{1}^{\prime}}^{\ast}%
(\mathbf{r})-c_{p_{1}^{\prime}}^{\ast}(\mathbf{r})\nabla_{\mathbf{r}}c_{p_{1}%
}(\mathbf{r}))
\]%
\[
+\sum_{p_{2}^{\prime}(\neq p_{2})}\frac{\langle c_{p_{2}^{\prime}%
}(1)|e\mathbf{E}\cdot\mathbf{r}_{1}|c_{p_{2}}(1)\rangle^{\ast}}{E_{c_{p_{2}%
^{\prime}}}-E_{c_{p_{2}}}}\int_{\Omega_{\mathbf{r}^{\prime}}}d\mathbf{r}%
(c_{p_{2}}(\mathbf{r})\nabla_{\mathbf{r}}c_{p_{2}^{\prime}}^{\ast}%
(\mathbf{r})-c_{p_{2}^{\prime}}^{\ast}(\mathbf{r})\nabla_{\mathbf{r}}c_{p_{2}%
}(\mathbf{r}))
\]%
\[
+\sum_{j_{1}^{\prime}(\neq j_{1})}\frac{\langle v_{j_{1}^{\prime}%
}(1)|e\mathbf{E}\cdot\mathbf{r}_{1}|v_{j_{1}}(1)\rangle}{E_{v_{j_{1}^{\prime}%
}}-E_{v_{j_{1}}}}\int_{\Omega_{\mathbf{r}^{\prime}}}d\mathbf{r}(v_{j_{1}%
^{\prime}}(\mathbf{r})\nabla_{\mathbf{r}}v_{j_{1}}^{\ast}(\mathbf{r}%
)-v_{j_{1}}^{\ast}(\mathbf{r})\nabla_{\mathbf{r}}v_{j_{1}^{\prime}}%
(\mathbf{r}))
\]%
\[
+\sum_{j_{2}^{\prime}(\neq j_{2})}\frac{\langle v_{j_{2}^{\prime}%
}(1)|e\mathbf{E}\cdot\mathbf{r}_{1}|v_{j_{2}}(1)\rangle}{E_{v_{j_{2}^{\prime}%
}}-E_{v_{j_{2}}}}\int_{\Omega}d\mathbf{r}(v_{j_{2}^{\prime}}(\mathbf{r}%
)\nabla_{\mathbf{r}}v_{j_{2}}^{\ast}(\mathbf{r})-v_{j_{2}}^{\ast}%
(\mathbf{r})\nabla_{\mathbf{r}}v_{j_{2}^{\prime}}(\mathbf{r}))
\]%
\[
+\sum_{p_{1}^{\prime}(\neq p_{1})}\frac{\langle c_{p_{1}^{\prime}%
}(1)|e\mathbf{E}\cdot\mathbf{r}_{1}|c_{p_{1}}(1)\rangle}{E_{c_{p_{1}^{\prime}%
}}-E_{c_{p_{1}}}}\int_{\Omega_{\mathbf{r}^{\prime}}}d\mathbf{r}(c_{p_{1}%
^{\prime}}(\mathbf{r})\nabla_{\mathbf{r}}c_{p_{1}}^{\ast}(\mathbf{r}%
)-c_{p_{1}}^{\ast}(\mathbf{r})\nabla_{\mathbf{r}}c_{p_{1}^{\prime}}%
(\mathbf{r}))
\]%
\begin{equation}
+\sum_{p_{2}^{\prime}(\neq p_{2})}\frac{\langle c_{p_{2}^{\prime}%
}(1)|e\mathbf{E}\cdot\mathbf{r}_{1}|c_{p_{2}}(1)\rangle}{E_{c_{p_{2}^{\prime}%
}}-E_{c_{p_{2}}}}\int_{\Omega_{\mathbf{r}^{\prime}}}d\mathbf{r}(c_{p_{2}%
^{\prime}}(\mathbf{r})\nabla_{\mathbf{r}}c_{p_{2}}^{\ast}(\mathbf{r}%
)-c_{p_{2}}^{\ast}(\mathbf{r})\nabla_{\mathbf{r}}c_{p_{2}^{\prime}}%
(\mathbf{r}))\} \label{2-liu}%
\end{equation}
Using Eq.(\ref{cdu}), one can pick off the conductivity from coupling between
2-electron excited states.

\subsection{Metals}

In a metal, the conduction band is partly filled. Relative to the Fermi
surface, holes and electrons are in the same conduction band. The energy
difference between hole and electron always can be taken as small. Beside the
non-existent energy gap, the difference between a metal and a semiconductor is
that the number of carriers $\thicksim\frac{k_{B}T}{E_{F}}N_{e}$ in the former
is greatly larger than that in the later. It is easy to check%
\begin{equation}
\int dr_{1}dr_{2}dr_{3}\frac{1}{\sqrt{3!}}\left\vert
\begin{array}
[c]{ccc}%
l_{1}^{\prime}(1) & l_{1}^{\prime}(2) & l_{1}^{\prime}(3)\\
l_{2}^{\prime}(1) & l_{2}^{\prime}(2) & l_{2}^{\prime}(3)\\
l_{3}^{\prime}(1) & l_{3}^{\prime}(2) & l_{3}^{\prime}(3)
\end{array}
\right\vert (r_{1}+r_{2}+r_{3})\frac{1}{\sqrt{3!}}\left\vert
\begin{array}
[c]{ccc}%
l_{1}(1) & l_{1}(2) & l_{1}(3)\\
l_{2}(1) & l_{2}(2) & l_{2}(3)\\
l_{3}(1) & l_{3}(2) & l_{3}(3)
\end{array}
\right\vert \label{33}%
\end{equation}%
\[
=\langle l_{1}^{\prime}|r|l_{1}\rangle(\delta_{l_{2}^{\prime}l_{2}}%
\delta_{l_{3}^{\prime}l_{3}}-\delta_{l_{2}^{\prime}l_{3}}\delta_{l_{3}%
^{\prime}l_{2}})+\langle l_{1}^{\prime}|r|l_{2}\rangle(\delta_{l_{2}^{\prime
}l_{3}}\delta_{l_{3}^{\prime}l_{1}}-\delta_{l_{2}^{\prime}l_{1}}\delta
_{l_{3}^{\prime}l_{3}})+\langle l_{1}^{\prime}|r|l_{3}\rangle(\delta
_{l_{2}^{\prime}l_{1}}\delta_{l_{3}^{\prime}l_{2}}-\delta_{l_{2}^{\prime}%
l_{2}}\delta_{l_{3}^{\prime}l_{1}})
\]%
\[
+\langle l_{2}^{\prime}|r|l_{1}\rangle(\delta_{l_{3}^{\prime}l_{2}}%
\delta_{l_{1}^{\prime}l_{3}}-\delta_{l_{3}^{\prime}l_{3}}\delta_{l_{1}%
^{\prime}l_{2}})+\langle l_{2}^{\prime}|r|l_{2}\rangle(\delta_{l_{3}^{\prime
}l_{3}}\delta_{l_{1}^{\prime}l_{1}}-\delta_{l_{3}^{\prime}l_{1}}\delta
_{l_{1}^{\prime}l_{3}})+\langle l_{2}^{\prime}|r|l_{3}\rangle(\delta
_{l_{3}^{\prime}l_{1}}\delta_{l_{1}^{\prime}l_{2}}-\delta_{l_{3}^{\prime}%
l_{2}}\delta_{l_{1}^{\prime}l_{1}})
\]%
\[
+\langle l_{3}^{\prime}|r|l_{1}\rangle(\delta_{l_{2}^{\prime}l_{3}}%
\delta_{l_{1}^{\prime}l_{2}}-\delta_{l_{2}^{\prime}l_{2}}\delta_{l_{1}%
^{\prime}l_{3}})+\langle l_{3}^{\prime}|r|l_{2}\rangle(\delta_{l_{2}^{\prime
}l_{1}}\delta_{l_{1}^{\prime}l_{3}}-\delta_{l_{2}^{\prime}l_{3}}\delta
_{l_{1}^{\prime}l_{1}})+\langle l_{3}^{\prime}|r|l_{3}\rangle(\delta
_{l_{2}^{\prime}l_{2}}\delta_{l_{1}^{\prime}l_{1}}-\delta_{l_{2}^{\prime}%
l_{1}}\delta_{l_{1}^{\prime}l_{2}})
\]
there are 9 terms, each term has 2 sub-terms (they form a determinant). By
induction method, one finds%
\begin{equation}
\langle\Lambda_{l_{1}^{\prime}l_{2}^{\prime}\cdots l_{N_{e}}^{\prime}%
}|-e\mathbf{E}\cdot\sum_{m=1}^{N_{e}}\mathbf{r}_{m}|\Lambda_{l_{1}l_{2}\cdots
l_{N_{e}}}\rangle=\sum_{j,k=1}^{N_{e}}(-)^{j+k}\langle\chi_{l_{j}^{\prime}%
}|-e\mathbf{E}\cdot\mathbf{r}|\chi_{l_{k}}\rangle D_{jk}^{(l_{1}l_{2}\cdots
l_{N_{e}};l_{1}^{\prime}l_{2}^{\prime}\cdots l_{N_{e}}^{\prime})} \label{smb}%
\end{equation}
where $D_{jk}^{(l_{1}l_{2}\cdots l_{N_{e}};l_{1}^{\prime}l_{2}^{\prime}\cdots
l_{N_{e}}^{\prime})}$ is a $(N_{e}-1)\times(N_{e}-1)$ determinant, each
element of $D$ is a Kronecker delta symbol. The row indices are ($l_{1}%
l_{2}\cdots l_{N_{e}}$) in which $l_{k}$ is removed, the column indices are
($l_{1}^{\prime}l_{2}^{\prime}\cdots l_{N_{e}}^{\prime}$) in which
$l_{j}^{\prime}$ is removed. For example%
\begin{equation}
D_{l_{2}l_{3}^{\prime}}^{(l_{1}l_{2}\cdots l_{N_{e}};l_{1}^{\prime}%
l_{2}^{\prime}\cdots l_{N_{e}}^{\prime})}=\left\vert
\begin{array}
[c]{ccccc}%
\delta_{l_{1}l_{1}^{\prime}} & \delta_{l_{1}l_{2}^{\prime}} & \delta
_{l_{1}l_{4}^{\prime}} & \cdots & \delta_{l_{1}l_{N_{e}}^{\prime}}\\
\delta_{l_{3}l_{1}^{\prime}} & \delta_{l_{3}l_{2}^{\prime}} & \delta
_{l_{3}l_{4}^{\prime}} & \cdots & \delta_{l_{3}l_{N_{e}}^{\prime}}\\
\delta_{l_{4}l_{1}^{\prime}} & \delta_{l_{4}l_{2}^{\prime}} & \delta
_{l_{4}l_{4}^{\prime}} &  & \delta_{l_{4}l_{N_{e}}^{\prime}}\\
\vdots &  &  &  & \vdots\\
\delta_{l_{N_{e}}l_{1}^{\prime}} & \delta_{l_{N_{e}}l_{2}^{\prime}} &
\delta_{l_{N_{e}}l_{4}^{\prime}} & \cdots & \delta_{l_{N_{e}}l_{N_{e}}%
^{\prime}}%
\end{array}
\right\vert \label{dm}%
\end{equation}
The first order change in the $N_{e}-$electron wave function is%
\begin{equation}
\Lambda_{l_{1}l_{2}\cdots l_{N_{e}}}^{\prime(1)}=\sum_{l_{1}^{\prime}%
l_{2}^{\prime}\cdots l_{N_{e}}^{\prime}}\sum_{jk}(-)^{j+k}\frac{\langle
\chi_{l_{j}^{\prime}}|-eE\cdot\mathbf{r}|\chi_{l_{k}}\rangle}{E_{l_{k}%
}-E_{l_{j}^{\prime}}}\Lambda_{l_{1}^{\prime}l_{2}^{\prime}\cdots l_{N_{e}%
}^{\prime}}D_{jk} \label{1wb}%
\end{equation}

Next, substitute Eq.(\ref{1wb}) into the expression of current density, one
finds the dc conductivity:%
\begin{equation}
\sigma_{\mu\nu}=\frac{e^{2}\hbar}{m\Omega}\sum_{l_{1}l_{2}\cdots l_{N_{e}}%
}W_{l_{1}l_{2}\cdots l_{N_{e}}}\operatorname{Im}\sum_{l_{1}^{\prime}%
l_{2}^{\prime}\cdots l_{N_{e}}^{\prime}}\sum_{j,k=1}^{N_{e}}(-)^{j+k}%
\frac{\langle\chi_{l_{j}^{\prime}}|x_{\nu}|\chi_{l_{k}}\rangle^{\ast}%
}{E_{l_{k}}-E_{l_{j}^{\prime}}}D_{jk}^{(l_{1}l_{2}\cdots l_{N_{e}}%
;l_{1}^{\prime}l_{2}^{\prime}\cdots l_{N_{e}}^{\prime})} \label{mco}%
\end{equation}%
\[
\sum_{p,q=1}^{N_{e}}(-)^{p+q}D_{pq}^{(l_{1}^{\prime}l_{2}^{\prime}\cdots
l_{N_{e}}^{\prime};l_{1}l_{2}\cdots l_{N_{e}})}\int d\mathbf{r}(\chi_{l_{p}%
}\frac{\partial\chi_{l_{q}^{\prime}}^{\ast}}{\partial x_{\mu}}-\chi
_{l_{q}^{\prime}}^{\ast}\frac{\partial\chi_{_{l_{p}}}}{\partial x_{\mu}})
\]
where $D_{pq}^{(l_{1}^{\prime}l_{2}^{\prime}\cdots l_{N_{e}}^{\prime}%
;l_{1}l_{2}\cdots l_{N_{e}})}$ is a $(N_{e}-1)\times(N_{e}-1)$ determinant,
each element of which is a Kronecker delta symbol. The row indices are
($l_{1}^{\prime}l_{2}^{\prime}\cdots l_{N_{e}}^{\prime}$) in which
$l_{q}^{\prime}$ is removed. The column indices are ($l_{1}l_{2}\cdots
l_{N_{e}}$) in which $l_{p}$ is removed.
\begin{equation}
W_{l_{1}l_{2}\cdots l_{N_{e}}}=%
{\displaystyle\prod\limits_{j=1}^{N_{e}}}
f(E_{l_{j}}),\text{ \ \ }f(E_{l_{j}})=\frac{1}{1+\exp(\frac{E_{l_{j}}-E_{F}%
}{k_{B}T})} \label{mw}%
\end{equation}
is the appearing probability of $N_{e}-$electron state $\Lambda_{l_{1}%
l_{2}\cdots l_{N_{e}}}$. Using the definition of Fermi distribution, it is
easy to check%
\begin{equation}
f(\varepsilon)=\frac{1}{e^{\beta\varepsilon}+1}=1-f(-\varepsilon),\text{
\ \ }\varepsilon=E-E_{F} \label{hol}%
\end{equation}
introducing an electron in a state above Fermi surface ($E>E_{F}$) is
equivalent to introduce a hole below Fermi surface ($E<E_{F}$). It is clear
from Eq.(\ref{mw}), the states around Fermi surface contribute most to
conductivity, as expected.

In a semiconductor, due to the energy gap $E_{g}$, the appearing probability
of a K-electron excited state includes a factor $e^{-KE_{g}/k_{B}T}$. To
calculate the conductivity, it is enough to restrict attention to the excited
states with few electrons. In a metal, the conduction band is half-filled, and
there exist many low-energy excited states. One must count all electrons
although only a shell $\frac{k_{B}T}{E_{F}}$ close to Fermi surface makes an
important contribution.

\subsection{Homogeneous doped semiconductor}

In a weakly n-type doped semiconductor, there are substitutional atoms or
interstitial atoms. The energy levels of the former lie just below the bottom
of conduction band; the energy levels of the later lie above the Fermi level
$\mu_{i}$ of the intrinsic matrix.\ There are three contributions to
conductivity: (1) electrons from substitutional donors, given by
Eq.(\ref{mco}), E$_{F}$ is the chemical potential of doped material in
Eq.(\ref{mw}); (2) carriers from interstitial atoms; (3) carriers from
intrinsic matrix. Parts (2) and (3) can be calculated by Eq.(\ref{ong}).
Similar consideration is applicable to weakly p-type doped semiconductors.

\section{Summary}

We discussed some foundational issues with respect to computing the
conductivity, and improved the Kubo-Greenwood formula by computing dc and ac
conductivity from current density, in which only the amplitude of probability,
and not the transition probability itself is used. Eqs. (\ref{cr}) and
(\ref{imc}) are key new contributions of this paper. In this method, the
expression of dc conductivity is extracted from the ac conductivity in a
direct way. We found that (1) the contribution from the states which are near
or in resonance with finite frequency external field is finite; (2) the
contribution from degenerate states in low frequency or zero frequency
external field is finite; (3) the energy conserving $\delta$-function does not
appear in the improved expression, thus one can avoid artificial numerical
broadening. In the formulation of many-body perturbation theory, i.e.
\textquotedblleft method (2)" for calculating the current density, the
many-electron statistics is displayed explicitly. One example is that the dc
conductivity of an intrinsic semiconductor at T=0K is zero. For an intrinsic
semiconductor, Kubo-Greenwood formula is the contribution from 1-electron
excited states.

\section{Appendices}

\subsection{Conductivity from degenerate states in a low frequency external
field}

We first compute the first order correction to $\chi_{p_{\mu}}^{\prime(0)}%
(t)$:%
\begin{equation}
\chi_{p_{\mu}}^{\prime(1)}(t)=\sum_{j}a_{p_{\mu}j}^{(1)}(t)\chi_{j}%
e^{-itE_{j}/\hbar}+\sum_{\mu^{\prime}}a_{p_{\mu}p_{\mu^{\prime}}}^{(1)}%
(t)\chi_{p_{\mu^{\prime}}}^{\prime(0)}e^{-itE_{p_{\mu^{\prime}}}/\hbar}
\label{1cd}%
\end{equation}
Now the zeroth order wave functions are%
\[
\cdots\chi_{k}\cdots;\text{ \ \ \ }\chi_{p_{1}}^{\prime(0)},\chi_{p_{2}%
}^{\prime(0)},\cdots,\chi_{p_{M}}^{\prime(0)}%
\]
At an initial moment, one electron is in state $\chi_{p_{\mu}}^{\prime(0)}$:
$a_{p_{\mu}}(-\infty)=1$ and other coefficients are zero. If the interaction
time with field is not too long, $a_{p_{\mu}}(t)$ is dominant. For a
non-degenerate state $\chi_{j}$, $a_{p_{\mu}j}^{(1)}(t)$ is determined by%
\begin{equation}
\frac{da_{j}(t)}{dt}=-\frac{i}{2\hbar}G_{jp_{\mu}}(e^{i\omega t}+e^{-i\omega
t})a_{p_{\mu}}(t)e^{i\omega_{jp_{\mu}}t} \label{nj}%
\end{equation}
The solution which satisfies initial condition $a_{j}(-\infty)=0$ is%
\begin{equation}
a_{p_{\mu}j}(t)=-\frac{1}{2\hbar}G_{jp_{\mu}}[\frac{e^{i(\omega+\omega
_{jp_{\mu}}+i\delta)t/\hbar}}{\omega_{jp_{\mu}}+\omega+i\delta}+\frac
{e^{i(\omega_{jp_{\mu}}-\omega+i\delta)t/\hbar}}{\omega_{jp_{\mu}}%
-\omega+i\delta}],\text{ \ }\delta\rightarrow0^{+} \label{sj}%
\end{equation}
For a member of the degenerate states $\chi_{p_{\mu^{\prime}}}$ ($\mu^{\prime
}\neq\mu$), $a_{p_{\mu^{\prime}}}(t)$ satisfies%
\begin{equation}
i\hbar\frac{da_{p_{\mu^{\prime}}}(t)}{dt}e^{-itE_{p_{\mu^{\prime}}}/\hbar
}=\sum_{k}a_{k}(t)G_{p_{\mu^{\prime}}k}e^{-itE_{k}/\hbar}\frac{e^{it\omega
}+e^{-it\omega}}{2} \label{dj}%
\end{equation}
and initial condition $a_{p_{\mu^{\prime}}}(-\infty)=0$. Index $k$ in RHS of
Eq.(\ref{dj}) runs over non-degenerate states only. Because Eq.(\ref{dign}),
no coupling among $\{\chi_{p_{\mu}}^{\prime(0)}\}$ in Eq.(\ref{dj}).
$a_{p_{\mu}p_{\mu^{\prime}}}^{(1)}(t)$ is given by%
\[
a_{p_{\mu}p_{\mu^{\prime}}}(t)=\frac{1}{4\hbar^{2}}\sum_{k}G_{p_{\mu^{\prime}%
}k}G_{kp_{\mu}}\{\frac{1}{\omega_{kp_{\mu}}+\omega}\frac{e^{it(\omega
_{p_{\mu^{\prime}}p_{\mu}}+2\omega+i\delta)}}{\omega_{p_{\mu^{\prime}}p_{\mu}%
}+2\omega+i\delta}+\frac{1}{\omega_{kp_{\mu}}+\omega}\frac{e^{it(\omega
_{p_{\mu^{\prime}}p_{\mu}}+i\delta)}}{\omega_{p_{\mu^{\prime}}p_{\mu}}%
+i\delta}%
\]%
\begin{equation}
+\frac{1}{\omega_{kp_{\mu}}-\omega}\frac{e^{it(\omega_{p_{\mu^{\prime}}p_{\mu
}}+i\delta)}}{\omega_{p_{\mu^{\prime}}p_{\mu}}+i\delta}+\frac{1}%
{\omega_{kp_{\mu}}-\omega}\frac{e^{it(\omega_{p_{\mu^{\prime}}p_{\mu}}%
-2\omega+i\delta)}}{\omega_{p_{\mu^{\prime}}p_{\mu}}-2\omega+i\delta}\}
\label{pmp}%
\end{equation}
When $\omega\rightarrow0$, all the denominators of Eqs.(\ref{sj}) and
(\ref{pmp}) are non-zero. For degenerate states in a zero frequency external
field, the artificial singularity of perturbation formula (\ref{adc}) is removed.

Combining Eqs.(\ref{1cd}), (\ref{sj}), (\ref{pmp}) and (\ref{acc}), the
macroscopic current density can be written as%
\begin{equation}
\mathbf{j}^{e}(\mathbf{r}^{\prime},t)=\mathbf{j}_{non}^{e}(\mathbf{r}^{\prime
},t)+\mathbf{j}_{c}^{e}(\mathbf{r}^{\prime},t)+\mathbf{j}_{s}^{e}%
(\mathbf{r}^{\prime},t)+\mathbf{j}_{c2}^{e}(\mathbf{r}^{\prime},t)+\mathbf{j}%
_{s2}^{e}(\mathbf{r}^{\prime},t)+\mathbf{j}_{0}^{e}(\mathbf{r}^{\prime})
\label{tj}%
\end{equation}
where $\mathbf{j}_{non}^{e}(\mathbf{r}^{\prime},t)$ is the contribution from
non-degenerate states, and is obtained by replacing $\sum_{\alpha=1}^{N_{e}}$
by $\sum_{\alpha=1}^{N_{e}-M}$ (sum over only non-degenerate states) in
Eq.(\ref{cvec}).%
\begin{equation}
\mathbf{j}_{c}^{e}(\mathbf{r}^{\prime},t)=\cos\omega t\frac{e}{2m\Omega
_{\mathbf{r}^{\prime}}}\sum_{l_{1}l_{2}\cdots l_{N_{e}}}W_{l_{1}l_{2}\cdots
l_{N_{e}}}\sum_{\mu=1}^{M}\sum_{j}(1-n_{j}) \label{c}%
\end{equation}%
\[
\lbrack\frac{1}{\omega_{jp_{\mu}}+\omega}+\frac{1}{\omega_{jp_{\mu}}-\omega
}]\operatorname{Im}G_{jp_{\mu}}\int_{\Omega_{\mathbf{r}^{\prime}}}%
d\mathbf{s}(\chi_{j}\nabla_{\mathbf{s}}\chi_{p_{\mu}}^{\prime(0)\ast}%
-\chi_{p_{\mu}}^{\prime(0)\ast}\nabla_{\mathbf{s}}\chi_{j})
\]
is the component with $\cos\omega t$ factor.%
\begin{equation}
\mathbf{j}_{s}^{e}(\mathbf{r}^{\prime},t)=\sin\omega t\frac{e}{2m\Omega
_{\mathbf{r}^{\prime}}}\sum_{l_{1}l_{2}\cdots l_{N_{e}}}W_{l_{1}l_{2}\cdots
l_{N_{e}}}\sum_{\mu=1}^{M}\sum_{j}(1-n_{j}) \label{s}%
\end{equation}%
\[
\lbrack\frac{1}{\omega_{jp_{\mu}}+\omega}-\frac{1}{\omega_{jp_{\mu}}-\omega
}]\operatorname{Re}G_{jp_{\mu}}\int_{\Omega_{\mathbf{r}^{\prime}}}%
d\mathbf{s}(\chi_{j}\nabla_{\mathbf{s}}\chi_{p_{\mu}}^{\prime(0)\ast}%
-\chi_{p_{\mu}}^{\prime(0)\ast}\nabla_{\mathbf{s}}\chi_{j})
\]
is the component with $\sin\omega t$ factor. $\mathbf{j}_{c}^{e}$ and
$\mathbf{j}_{s}^{e}$ come from coupling non-degenerate states with the
degenerate manifold.%
\begin{equation}
\mathbf{j}_{c2}^{e}(\mathbf{r}^{\prime},t)=\cos2\omega t\frac{e}{4\hbar
m\Omega_{\mathbf{r}^{\prime}}}\sum_{l_{1}l_{2}\cdots l_{N_{e}}}W_{l_{1}%
l_{2}\cdots l_{N_{e}}}\sum_{\mu=1}^{M}\sum_{\mu^{\prime}}(1-n_{\mu^{\prime}%
})\sum_{k} \label{c2}%
\end{equation}%
\[
(\frac{1}{\omega_{kp_{\mu}}+\omega}\frac{1}{\omega_{p_{\mu^{\prime}}p_{\mu}%
}+2\omega}+\frac{1}{\omega_{kp_{\mu}}-\omega}\frac{1}{\omega_{p_{\mu^{\prime}%
}p_{\mu}}-2\omega})\operatorname{Im}G_{p_{\mu^{\prime}}k}G_{kp_{\mu}}%
\int_{\Omega_{\mathbf{r}^{\prime}}}d\mathbf{s}(\chi_{p_{\mu}}^{\prime(0)\ast
}\nabla_{\mathbf{s}}\chi_{p_{\mu^{\prime}}}^{\prime(0)}-\chi_{p_{\mu^{\prime}%
}}^{\prime(0)}\nabla_{\mathbf{s}}\chi_{p_{\mu}}^{\prime(0)\ast})]
\]
is the component with $\cos2\omega t$ factor.%
\begin{equation}
\mathbf{j}_{s2}^{e}(\mathbf{r}^{\prime},t)=\sin2\omega t\frac{e}{4\hbar
m\Omega_{\mathbf{r}^{\prime}}}\sum_{l_{1}l_{2}\cdots l_{N_{e}}}W_{l_{1}%
l_{2}\cdots l_{N_{e}}}\sum_{\mu=1}^{M}\sum_{\mu^{\prime}}(1-n_{\mu^{\prime}%
})\sum_{k} \label{s2}%
\end{equation}%
\[
(\frac{1}{\omega_{kp_{\mu}}-\omega}\frac{1}{\omega_{p_{\mu^{\prime}}p_{\mu}%
}-2\omega}-\frac{1}{\omega_{kp_{\mu}}+\omega}\frac{1}{\omega_{p_{\mu^{\prime}%
}p_{\mu}}+2\omega})\operatorname{Re}G_{p_{\mu^{\prime}}k}G_{kp_{\mu}}%
\int_{\Omega_{\mathbf{r}^{\prime}}}d\mathbf{s}(\chi_{p_{\mu^{\prime}}}%
^{\prime(0)}\nabla_{\mathbf{s}}\chi_{p_{\mu}}^{\prime(0)\ast}-\chi_{p_{\mu}%
}^{\prime(0)\ast}\nabla_{\mathbf{s}}\chi_{p_{\mu^{\prime}}}^{\prime(0)})
\]
is the component with $\sin2\omega t$ factor.%
\begin{equation}
\mathbf{j}_{0}^{e}(\mathbf{r}^{\prime})=\frac{e}{4\hbar m\Omega_{\mathbf{r}%
^{\prime}}}\sum_{l_{1}l_{2}\cdots l_{N_{e}}}W_{l_{1}l_{2}\cdots l_{N_{e}}}%
\sum_{\mu=1}^{M}\sum_{\mu^{\prime}}(1-n_{\mu^{\prime}})\sum_{k}(\frac
{1}{\omega_{kp_{\mu}}+\omega}\frac{1}{\omega_{p_{\mu^{\prime}}p_{\mu}}}%
+\frac{1}{\omega_{kp_{\mu}}-\omega}\frac{1}{\omega_{p_{\mu^{\prime}}p_{\mu}}})
\label{0}%
\end{equation}%
\[
\operatorname{Im}G_{p_{\mu^{\prime}}k}G_{kp_{\mu}}\int_{\Omega_{\mathbf{r}%
^{\prime}}}d\mathbf{s}(\chi_{p_{\mu}}^{\prime(0)\ast}\nabla_{\mathbf{s}}%
\chi_{p_{\mu^{\prime}}}^{\prime(0)}-\chi_{p_{\mu^{\prime}}}^{\prime(0)}%
\nabla_{\mathbf{s}}\chi_{p_{\mu}}^{\prime(0)\ast})]
\]
is the component without time variation factor. Using Eq.(\ref{cdu}), one can
easily identify the conductivity.

In Eqs.(\ref{c})-(\ref{0}), the summation over $j$ or $k$ is not restricted to
($l_{1}l_{2}\cdots l_{N_{e}}$): it extends to all single-electron states.
$\mathbf{j}_{c2}^{e}$, $\mathbf{j}_{s2}^{e}$ and $\mathbf{j}_{0}^{e}$ come
from indirect coupling among $\{\chi_{p_{\mu}}^{\prime(0)},$ $\mu
=1,2,\cdots,M\}$ through non-degenerate states. Interaction with external
field appear twice in Eq.(\ref{pmp}), three new time factors $\cos2\omega t$,
$\sin2\omega t$ and 1, which are different from the original time factors
$e^{-it\omega}$ and $e^{it\omega}$ of the external field, come from the
2$^{\text{nd}}$ order harmonic generations.

\subsection{Conductivity from resonant states}

To compute the contribution of two groups of resonant states to the current,
we need the 1$^{\text{st}}$ order wave function%
\begin{equation}
\chi_{q}^{\prime(1)}(t)=\sum_{s}a_{s}^{q(1)}e^{-iE_{s}t/\hbar}\chi_{s}%
+\sum_{k}a_{n_{k}}^{q(1)}e^{-iE_{n}t/\hbar}\chi_{n_{k}}+\sum_{j}a_{m_{j}%
}^{q(1)}e^{-iE_{m}t/\hbar}\chi_{m_{j}} \label{1jr}%
\end{equation}
where $s$ indicates the states which do not belong to the upper and lower
degenerate groups. $j$ scans over the upper group ($\chi_{m_{1}},\chi_{m_{2}%
},\cdots,\chi_{m_{M}}$), $k$ scans over the lower group ($\chi_{n_{1}}%
,\chi_{n_{2}},\cdots,\chi_{n_{M^{\prime}}}$). Suppose initially the system is
in the $q^{\text{th}}$ mode of the resonance states, the first order evolution
equation is then:%
\[
i\hbar\frac{da_{s}^{(1)}}{dt}=\sum_{j}[F_{sm_{j}}e^{it(\omega_{sm_{j}}%
-\omega)}+F_{m_{j}s}^{\ast}e^{it(\omega_{sm_{j}}+\omega)}]a_{m_{j}}^{(0)}%
\]%
\begin{equation}
+\sum_{k}[F_{sn_{k}}e^{it(\omega_{sn_{k}}-\omega)}+F_{n_{k}s}^{\ast
}e^{it(\omega_{sn_{k}}+\omega)}]a_{n_{k}}^{(0)} \label{gzs}%
\end{equation}
The solution of Eq.(\ref{gzs}) is%
\begin{equation}
a_{s}^{(1)}(t)=-\frac{1}{\hbar}\sum_{j}[F_{sm_{j}}\frac{e^{it(\omega_{sm_{j}%
}-\omega+\alpha_{q})}}{\omega_{sm_{j}}-\omega+\alpha_{q}}+F_{m_{j}s}^{\ast
}\frac{e^{it(\omega_{sm_{j}}+\omega+\alpha_{q})}}{\omega_{sm_{j}}%
+\omega+\alpha_{q}}]a_{m_{j}}^{q0} \label{sjie}%
\end{equation}%
\[
-\frac{1}{\hbar}\sum_{k}[F_{sn_{k}}\frac{e^{it(\omega_{sn_{k}}-\omega
+\alpha_{q}-\epsilon)}}{\omega_{sn_{k}}-\omega+\alpha_{q}-\epsilon}+F_{n_{k}%
s}^{\ast}\frac{e^{it(\omega_{sn_{k}}+\omega+\alpha_{q}-\epsilon)}}%
{\omega_{sn_{k}}+\omega+\alpha_{q}-\epsilon}]b_{n_{k}}^{q0}%
\]
For a member of the upper group, the first order probability amplitude is
determined by:%
\begin{equation}
i\hbar\frac{da_{m_{j}}^{(1)}}{dt}=\sum_{k=1}^{M^{\prime}}F_{n_{k}m_{j}}^{\ast
}e^{it(2\omega+\epsilon)}a_{n_{k}}^{(0)}+\sum_{j^{\prime}(\neq j)}%
[F_{m_{j}m_{j^{\prime}}}e^{-it\omega}+F_{m_{j^{\prime}}m_{j}}^{\ast
}e^{it\omega}]a_{m_{j^{\prime}}}^{(0)} \label{mgz1}%
\end{equation}%
\[
+\sum_{s}[F_{m_{j}s}e^{it(\omega_{m_{j}s}-\omega)}+F_{sm_{j}}^{\ast
}e^{it(\omega_{m_{j}s}+\omega)}]a_{s}^{(1)},\text{ \ \ \ \ \ }j=1,2,\cdots,M
\]
Using Eq.(\ref{sjie}), the solution of eq.(\ref{mgz1}) is%
\begin{equation}
a_{m_{j}}^{(1)}=-\frac{1}{\hbar}\sum_{k=1}^{M^{\prime}}F_{n_{k}m_{j}}^{\ast
}\frac{e^{it(2\omega+\alpha_{q})}}{2\omega+\alpha_{q}}b_{n_{k}}^{q0}-\frac
{1}{\hbar}\sum_{j^{\prime}(\neq j)}[F_{m_{j}m_{j^{\prime}}}\frac
{e^{it(\alpha_{q}-\omega)}}{\alpha_{q}-\omega}+F_{m_{j^{\prime}}m_{j}}^{\ast
}\frac{e^{it(\alpha_{q}+\omega)}}{\alpha_{q}+\omega}]a_{m_{j^{\prime}}}^{q0}
\label{jso}%
\end{equation}%
\[
+\frac{1}{\hbar^{2}}\sum_{s}F_{m_{j}s}\{\sum_{j^{\prime}}[F_{sm_{j^{\prime}}%
}\frac{e^{i(\alpha_{q}-2\omega)t}}{(\omega_{sm_{j^{\prime}}}-\omega+\alpha
_{q})(\alpha_{q}-2\omega)}+F_{m_{j}s}^{\ast}\frac{e^{it\alpha_{q}}}%
{(\omega_{sm_{j^{\prime}}}+\omega+\alpha_{q})\alpha_{q}}]a_{m_{j^{\prime}}%
}^{q0}%
\]%
\[
+\sum_{k}[F_{sn_{k}}\frac{e^{i(\alpha_{q}-\omega)t}}{(\omega_{sn_{k}}%
-\omega+\alpha_{q}-\epsilon)(\alpha_{q}-\omega)}+F_{n_{k}s}^{\ast}%
\frac{e^{i(\alpha_{q}+\omega)t}}{(\omega_{sn_{k}}+\omega+\alpha_{q}%
-\epsilon)(\alpha_{q}+\omega)}]b_{n_{k}}^{q0}\}
\]%
\[
+\frac{1}{\hbar^{2}}\sum_{s}F_{sm_{j}}^{\ast}\{\sum_{j^{\prime}}%
[F_{sm_{j^{\prime}}}\frac{e^{it\alpha_{q}}}{(\omega_{sm_{j^{\prime}}}%
-\omega+\alpha_{q})\alpha_{q}}+F_{m_{j^{\prime}}s}^{\ast}\frac{e^{i(\alpha
_{q}+2\omega)t}}{(\omega_{sm_{j^{\prime}}}+\omega+\alpha_{q})(\alpha
_{q}+2\omega)}]a_{m_{j}^{\prime}}^{q0}%
\]%
\[
+\sum_{k}[F_{sn_{k}}\frac{e^{i(\alpha_{q}+\omega)t}}{(\omega_{sn_{k}}%
-\omega+\alpha_{q}-\epsilon)(\alpha_{q}+\omega)}+F_{n_{k}s}^{\ast}%
\frac{e^{it(3\omega+\alpha_{q})}}{(\omega_{sn_{k}}+\omega+\alpha_{q}%
-\epsilon)(3\omega+\alpha_{q})}]b_{n_{k}}^{q0}\}
\]
For a member of the lower group, the first order probability amplitude is
determined by:%
\[
i\hbar\frac{da_{n_{k}}^{(1)}}{dt}=\sum_{j=1}^{M}F_{n_{k}m_{j}}e^{-it(2\omega
+\epsilon)}a_{m_{j}}^{(0)}+\sum_{k^{\prime}(\neq k)}[F_{n_{k}n_{k^{\prime}}%
}e^{-it\omega}+F_{n_{k^{\prime}}n_{k}}^{\ast}e^{it\omega}]a_{n_{k^{\prime}}%
}^{(0)}%
\]%
\begin{equation}
+\sum_{s}[F_{n_{k}s}e^{it(\omega_{n_{k}s}-\omega)}+F_{sn_{k}}^{\ast
}e^{it(\omega_{n_{k}s}+\omega)}]a_{s}^{(1)},\text{ \ \ \ \ \ }k=1,2,\cdots
,M^{\prime} \label{kz1}%
\end{equation}
Using Eq.(\ref{sjie}), the solution of Eq.(\ref{kz1}) is%
\begin{equation}
a_{n_{k}}^{(1)}(t)=-\frac{1}{\hbar}\sum_{j=1}^{M}F_{n_{k}m_{j}}\frac
{e^{it(\alpha_{q}-2\omega-\epsilon)}}{\alpha_{q}-2\omega-\epsilon}a_{m_{j}%
}^{q0}-\frac{1}{\hbar}\sum_{k^{\prime}(\neq k)}[F_{n_{k}n_{k^{\prime}}}%
\frac{e^{it(\alpha_{q}-\omega-\epsilon)}}{\alpha_{q}-\omega-\epsilon
}+F_{n_{k^{\prime}}n_{k}}^{\ast}\frac{e^{it(\alpha_{q}+\omega-\epsilon)}%
}{\alpha_{q}+\omega-\epsilon}]b_{n_{k^{\prime}}}^{q0} \label{kjie}%
\end{equation}%
\[
+\frac{1}{\hbar^{2}}\sum_{s}F_{n_{k}s}\{\sum_{j}[F_{sm_{j}}\frac
{e^{it(\alpha_{q}-3\omega-\epsilon)}}{(\omega_{sm_{j}}-\omega+\alpha
_{q})(\alpha_{q}-3\omega-\epsilon)}+F_{m_{j}s}^{\ast}\frac{e^{it(\alpha
_{q}-\omega-\epsilon)}}{(\omega_{sm_{j}}+\omega+\alpha_{q})(\alpha_{q}%
-\omega-\epsilon)}]a_{m_{j}}^{q0}%
\]%
\[
+\sum_{k^{\prime}}[F_{sn_{k^{\prime}}}\frac{e^{it(\alpha_{q}-2\omega
-\epsilon)}}{(\omega_{sn_{k^{\prime}}}-\omega+\alpha_{q}-\epsilon)(\alpha
_{q}-2\omega-\epsilon)}+F_{n_{k^{\prime}}s}^{\ast}\frac{e^{it(\alpha
_{q}-\epsilon)}}{(\omega_{sn_{k^{\prime}}}+\omega+\alpha_{q}-\epsilon
)(\alpha_{q}-\epsilon)}]b_{n_{k^{\prime}}}^{q0}\}
\]%
\[
+\frac{1}{\hbar^{2}}\sum_{s}F_{sn_{k}}^{\ast}\{\sum_{j}[F_{sm_{j}}%
\frac{e^{it(\alpha_{q}-\omega-\epsilon)}}{(\omega_{sm_{j}}-\omega+\alpha
_{q})(\alpha_{q}-\omega-\epsilon)}+F_{m_{j}s}^{\ast}\frac{e^{it(\alpha
_{q}+\omega-\epsilon)}}{(\omega_{sm_{j}}+\omega+\alpha_{q})(\alpha_{q}%
+\omega-\epsilon)}]a_{m_{j}}^{q0}%
\]%
\[
+\sum_{k^{\prime}}[F_{sn_{k^{\prime}}}\frac{e^{it(\alpha_{q}-\epsilon)}%
}{(\omega_{sn_{k^{\prime}}}-\omega+\alpha_{q}-\epsilon)(\alpha_{q}-\epsilon
)}+F_{n_{k^{\prime}}s}^{\ast}\frac{e^{it(\alpha_{q}+2\omega-\epsilon)}%
}{(\omega_{sn_{k^{\prime}}}+\omega+\alpha_{q}-\epsilon)(\alpha_{q}%
+2\omega-\epsilon)}]b_{n_{k^{\prime}}}^{q0}\}
\]

Substituting Eqs.(\ref{sjie}), (\ref{jso}) and (\ref{kjie}) into Eqs.
(\ref{1jr}), by means of Eq.(\ref{acc}), one can find current density. The
full formula is too long to write out, we only write down the contribution
from coupling between non-resonant states and resonant states: in
Eq.(\ref{1jr}) only keep the first term. The current density with time factor
$\cos\omega t$ is%
\begin{equation}
\mathbf{j}_{c}(\mathbf{r},t)=\frac{e}{m\Omega_{\mathbf{r}}}\cos\omega
t\int_{\Omega_{\mathbf{r}}}d\mathbf{s}\sum_{l_{1}l_{2}\cdots l_{N_{e}}%
}W_{l_{1}l_{2}\cdots l_{N_{e}}}\sum_{qs}(1-n_{s}) \label{rco}%
\end{equation}%
\[
\operatorname{Im}\{\sum_{jj^{\prime}}a_{m_{j}}^{q0}a_{m_{j^{\prime}}}^{q0\ast
}(\chi_{m_{j}}\nabla\chi_{s}^{\ast}-\chi_{s}^{\ast}\nabla\chi_{m_{j}%
})[F_{sm_{j^{\prime}}}^{\ast}\frac{1}{\omega_{sm_{j^{\prime}}}-\omega
+\alpha_{q}}+F_{m_{j^{\prime}}s}\frac{1}{\omega_{sm_{j^{\prime}}}%
+\omega+\alpha_{q}}]
\]%
\[
+\sum_{kk^{\prime}}b_{n_{k}}^{q0}b_{n_{k^{\prime}}}^{q0\ast}(\chi_{n_{k}%
}\nabla\chi_{s}^{\ast}-\chi_{s}^{\ast}\nabla\chi_{n_{k}})[F_{sn_{k^{\prime}}%
}^{\ast}\frac{1}{\omega_{sn_{k^{\prime}}}-\omega+\alpha_{q}-\epsilon
}+F_{n_{k^{\prime}}s}\frac{1}{\omega_{sn_{k^{\prime}}}+\omega+\alpha
_{q}-\epsilon}]\}
\]
The current density with time factor $\sin\omega t$ is%

\begin{equation}
\mathbf{j}_{s}(\mathbf{r},t)=\frac{-ie}{m\Omega_{\mathbf{r}}}\sin\omega
t\int_{\Omega_{\mathbf{r}}}d\mathbf{s}\sum_{l_{1}l_{2}\cdots l_{N_{e}}%
}W_{l_{1}l_{2}\cdots l_{N_{e}}}\sum_{qs}(1-n_{s}) \label{rsin}%
\end{equation}%
\[
\operatorname{Im}[\sum_{jj^{\prime}}a_{m_{j}}^{q0\ast}a_{m_{j^{\prime}}}%
^{q0}(\chi_{s}\nabla\chi_{m_{j}}^{\ast}+\chi_{m_{j}}^{\ast}\nabla\chi
_{s})[F_{sm_{j^{\prime}}}\frac{1}{\omega_{sm_{j^{\prime}}}-\omega+\alpha_{q}%
}-F_{m_{j^{\prime}}s}^{\ast}\frac{1}{\omega_{sm_{j^{\prime}}}+\omega
+\alpha_{q}}]
\]%
\[
+\sum_{kk^{\prime}}b_{n_{k}}^{q0\ast}b_{n_{k^{\prime}}}^{q0}(\chi_{s}%
\nabla\chi_{n_{k}}^{\ast}+\chi_{n_{k}}^{\ast}\nabla\chi_{s})[F_{sn_{k^{\prime
}}}\frac{1}{\omega_{sn_{k^{\prime}}}-\omega+\alpha_{q}-\epsilon}%
-F_{n_{k^{\prime}}s}^{\ast}\frac{1}{\omega_{sn_{k^{\prime}}}+\omega+\alpha
_{q}-\epsilon}]
\]
The current density with time factor $\cos2\omega t$ is%

\begin{equation}
\mathbf{j}_{c2}(\mathbf{r},t)=\frac{-ie}{m\Omega_{\mathbf{r}}}\cos2\omega
t\int_{\Omega_{\mathbf{r}}}d\mathbf{s}\sum_{l_{1}l_{2}\cdots l_{N_{e}}%
}W_{l_{1}l_{2}\cdots l_{N_{e}}}\sum_{qs}(1-n_{s})\sum_{jk} \label{rc2}%
\end{equation}%
\[
\operatorname{Re}[a_{m_{j}}^{q0\ast}b_{n_{k}}^{q0}(\chi_{s}\nabla\chi_{m_{j}%
}^{\ast}+\chi_{m_{j}}^{\ast}\nabla\chi_{s})F_{n_{k}s}^{\ast}\frac{1}%
{\omega_{sn_{k}}+\omega+\alpha_{q}-\epsilon}%
\]%
\[
+a_{m_{j}}^{q0}b_{n_{k}}^{q0\ast}(\chi_{n_{k}}^{\ast}\nabla\chi_{s}+\chi
_{s}\nabla\chi_{n_{k}}^{\ast})F_{sm_{j}}\frac{1}{\omega_{sm_{j}}-\omega
+\alpha_{q}}]
\]
The current density with time factor $\sin2\omega t$ is%

\begin{equation}
\mathbf{j}_{s2}(\mathbf{r},t)=\frac{-ie}{m\Omega_{\mathbf{r}}}\sin2\omega
t\int_{\Omega_{\mathbf{r}}}d\mathbf{s}\sum_{l_{1}l_{2}\cdots l_{N_{e}}%
}W_{l_{1}l_{2}\cdots l_{N_{e}}}\sum_{qs}(1-n_{s})\sum_{jk} \label{rs2}%
\end{equation}%
\[
\operatorname{Im}[a_{m_{j}}^{q0}b_{n_{k}}^{q0\ast}(\chi_{m_{j}}\nabla\chi
_{s}^{\ast}+\chi_{s}^{\ast}\nabla\chi_{m_{j}})F_{n_{k}s}\frac{1}%
{\omega_{sn_{k}}+\omega+\alpha_{q}-\epsilon}%
\]%
\[
+a_{m_{j}}^{q0}b_{n_{k}}^{q0\ast}(\chi_{n_{k}}^{\ast}\nabla\chi_{s}+\chi
_{s}\nabla\chi_{n_{k}}^{\ast})F_{sm_{j}}\frac{1}{\omega_{sm_{j}}-\omega
+\alpha_{q}}]
\]
The current density without time variation factor is%
\begin{equation}
\mathbf{j}_{0}(\mathbf{r})=\frac{-ie}{m\Omega_{\mathbf{r}}}\int_{\Omega
_{\mathbf{r}}}d\mathbf{s}\sum_{l_{1}l_{2}\cdots l_{N_{e}}}W_{l_{1}l_{2}\cdots
l_{N_{e}}}\sum_{qs}(1-n_{s})\sum_{jk} \label{r0}%
\end{equation}%
\[
\operatorname{Re}[a_{m_{j}}^{q0}b_{n_{k}}^{q0\ast}(\chi_{m_{j}}\nabla\chi
_{s}^{\ast}+\chi_{s}^{\ast}\nabla\chi_{m_{j}})F_{sn_{k}}^{\ast}\frac{1}%
{\omega_{sn_{k}}-\omega+\alpha_{q}-\epsilon}%
\]%
\[
+a_{m_{j}}^{q0\ast}b_{n_{k}}^{q0}(\chi_{n_{k}}\nabla\chi_{s}^{\ast}+\chi
_{s}^{\ast}\nabla\chi_{n_{k}})F_{m_{j}s}\frac{1}{\omega_{sm_{j}}+\omega
+\alpha_{q}}]
\]
The contribution to current from two groups of resonant states is finite.
Using Eq.(\ref{cdu}), one can again read off conductivity.

\section{Acknowledgements}

We thank the Army Research Office for support under MURI W91NF-06-2-0026, and
the National Science Foundation for support under grants DMR 0600073 and
0605890. DAD thanks the Leverhulme Trust (UK) and the National Science
Foundation for sabbatical support.


\begin{thebibliography}{99}                                                                                               %


\bibitem {ale}P. B. Allen and J. Q. Broughton, J. Phys. Chem. \textbf{91},
4964 (1987).

\bibitem {gali}G. Galli, R. M. Martin, R. Car and M. Parrinello, Phys. Rev. B
\textbf{42}, 7470 (1990).

\bibitem {abt}T. A. Abtew, M. Zhang and D. A. Drabold, Phys. Rev.
B\textbf{76}, 045212 (2007).

\bibitem {heli}W. Lorenzen, B. Holst and R. Redmer, Phys. Rev. Lett.
\textbf{102}, 115701 (2009).

\bibitem {low}S. Lowitzer, D. K\"{o}dderitzsch, H. Ebert, and J. B. Staunton,
Phys. Rev. B \textbf{79}, 115109 (2009).

\bibitem {cle}J. Cl\'{e}rouin, P. Noiret, V. N. Korobenko and A. D. Rakhel,
Phys. Rev. B \textbf{78}, 224203 (2008).

\bibitem {Gre}D. A. Greenwood, Proc. Phys. Soc. (London) \textbf{71}, 585 (1958).

\bibitem {md}N. F. Mott and E. A. Davis, Electronic Processes in
Non-crystalline Materials, Clarendon Press, Oxford (1971).

\bibitem {mos}L. L. Moseley and T. Lukes, Am. J. Phys. \textbf{46}, 676 (1978).

\bibitem {Over}H. Overhof and P. Thomas \textit{Electronic Transport in
Hydrogenated Amorphous Silicon} Springer Tracts in Modern Physics No. 114
(Springer, Berlin, 1989).

\bibitem {san}D. S\'{a}nchez-Portal, P. Ordej\'{o}n, and E. Canadell,
Structure and Bonding \textbf{113}, 103-170 (2004).

\bibitem {ll}L. D. Landua and E.M. Lifshitz, Qunatum Mechanics, 3rd edition,
Pergamon Press, Oxford (1977).

\bibitem {robin}F. N. H. Robinson, Macroscopic electromagnetism, Pergamon
Press, Oxford (1973).

\bibitem {jac}J. D. Jackson, Classical electrodynamics, 3rd edition, Wiley,
New York (1999).

\bibitem {kub}R. Kubo, J. Phys. Soc. Jpn. \textbf{12}, 570 (1957).

\bibitem {lut}J. M. Luttinger, Physical Review \textbf{135}, A1505, (1964).

\bibitem {her}C. Herring, Physical Review \textbf{52}, 365 (1937).

\bibitem {ott}R. J. Elliott, Phys. Rev. \textbf{96}, 280, (1954).

\bibitem {cov}L. M. Falicov, Group Theory and Its Physcial Applications, Univ.
of Chicago Press, Chicago, Illinois (1966).

\bibitem {ash}N. W. Ashcroft and N. D. Mermin, Solid State Physics, Holt,
Rinehart and Winston, NewYork (1976).
\end{thebibliography}
\end{document}